\documentclass[twocolumn,showpacs,preprintnumbers,amsmath,amssymb]{revtex4}

\usepackage{graphicx}
\usepackage{dcolumn}
\usepackage{bm}
\usepackage{subfigure}

\allowdisplaybreaks 

\begin{document}

\preprint{APS/123-QED}

\title{Helical turbulent Prandtl number in the $A$ model of passive advection: Two loop approximation}

\author{M.\,Hnati\v{c}}
\affiliation{Faculty of Sciences, P.J. Safarik University, Ko\v{s}ice, Slovakia}
\affiliation{Institute of Experimental Physics, SAS,
         Watsonova 47, 040 01 Ko\v{s}ice, Slovakia
        }
\affiliation{Bogoliubov Laboratory of Theoretical Physics, Joint Institute for Nuclear Research,
        141 980 Dubna, Moscow Region, Russian Federation
        }
        
\author{P.\,Zalom}
\affiliation{Institute of Experimental Physics, SAS,
         Watsonova 47, 040 01 Ko\v{s}ice, Slovakia
        }
\affiliation{Bogoliubov Laboratory of Theoretical Physics, Joint Institute for Nuclear Research,
        141 980 Dubna, Moscow Region, Russian Federation
        }
\draft

\date{\today}

\begin{abstract}
Using the field theoretic renormalization group technique in the two-loop approximation, turbulent Prandtl numbers are obtained in the general $A$ model of passive vector advected by fully developed turbulent velocity field with violation of spatial parity introduced via continuous parameter $\rho$ ranging from $\rho=0$ (no violation of spatial parity) to $|\rho|=1$ (maximum violation of spatial parity). Values of $A$ represent a continuously adjustable parameter which governs the interaction structure of the model. In non-helical environments, we demonstrate that $A$ is however restricted to the interval $-1.723 \leq A \leq 2.800$ (rounded on the last presented digit) due to the constraints of two-loop calculations. However, when $\rho >0.749$ (rounded on the last presented digit) restrictions may be removed. Furthermore, three physically important cases $A \in \{-1, 0, 1\}$ are shown to lie deep within the allowed interval of $A$ for all values of $\rho$. For the model of linearized Navier-Stokes equations ($A = -1$) up to date unknown helical values of turbulent Prandtl number have been shown to equal $1$ regardless of parity violation. Furthermore, we have shown that interaction parameter $A$ exerts strong influence on advection diffusion processes in turbulent environments with broken spatial parity. In explicit, depending on actual value of $A$ turbulent Prandtl number may increase or decrease with $\rho$. By varying $A$ continuously we explain high stability of kinematic MHD model ($A=1$) against helical effects as a result of its closeness to $A = 0.912$ (rounded on the last presented digit) case where helical effects are completely suppressed. Contrary, for the physically important $A=0$ model we show that it lies deep within the interval of models where helical effects cause the turbulent Prandtl number to decrease with $|\rho|$. We thus identify internal structure of interactions given by parameter $A$, and not the vector character of the admixture itself to be the dominant factor influencing diffusion advection processes in the helical $A$ model which significantly refines the conclusions of Ref.\cite{Jurcisin2014}.
\end{abstract}

\maketitle

\section{\label{sec1} Introduction}

Diffusion advection processes in turbulent environments represent both experimentally and theoretically important topic of study in the field of fluid motion \cite{Yoshizawa,Biskamp,MoninBook,McComb,Shraiman}. In this respect, the so-called Prandtl number is frequently used to compactly characterize the quantitative properties of flows under the study \cite{Biskamp,MoninBook}. For all admixture types, it is defined as the dimensionless ratio of the coefficient of kinematic viscosity to the corresponding diffusion coefficient of given admixture. For example in the case of thermal diffusivity, the corresponding (scalar) Prandtl number equals to the ratio of kinematic viscosity to the coefficient of molecular diffusivity \cite{MoninBook}. Since both the kinematic viscosity and the diffusion coefficient for given admixture are material and flow specific quantities the resulting Prandtl numbers have always to be specified at distinct conditions required to characterize the flow and are thus often found in property tables alongside other material specific properties \cite{Yoshizawa,Biskamp,Coulson,Chua}.

However, in the high Reynolds number limit the state of fully developed turbulence manifests itself by reaching effective material and flow independent values for both the kinematic viscosity and the corresponding diffusion coefficient. We commonly refer to such effective values as the turbulent viscosity coefficient and turbulent diffusion coefficient \cite{MoninBook,McComb}. Consequently, in fully developed turbulent flows the resulting values of Prandtl numbers are universal for given admixture and do not depend on microscopic nor macroscopic properties of the flow under the consideration. Usually, we refer to them as turbulent Prandtl numbers of given admixture type \cite{Yoshizawa,Biskamp,Chang,VasilevBook}. 

In other words, the state of fully developed turbulence allows for studying of advection diffusion processes on a general material and  flow unbiased manner \cite{MoninBook,McComb}. Moreover, it is well known that fully developed turbulent systems are well tractable for analytic investigations which would otherwise be difficult or even impossible \cite{VasilevBook,AdzhemyanBook}. Fully developed turbulent flows represent thus theoretically as well as experimentally valuable scenario for analytic studies of how different admixtures are transported within the underlying turbulent environment.

In this respect, several authors have recently analyzed the question of how tensorial nature of admixtures
under the consideration may alter the diffusion advection processes, see for example Refs.\,\cite{Jurcisin2014,Antonov2015,Antonov2015a,Jurcisin2016,Adzhemyan2005} for more details. As a starting point for the present analysis, we discuss briefly Refs.\,\cite{Jurcisin2014,Adzhemyan2005} where the aforementioned turbulent Prandtl number have been used to approach the problem. In Ref.\,\cite{Adzhemyan2005}, turbulent scalar Prandtl number has been investigated in the model of passive advection while in Ref.\,\cite{Jurcisin2014} two other models, namely the so called kinematic MHD model and a passively advected vector field within the $A = 0$ model, have been included into the analysis. As argued by authors of Ref.\,\cite{Jurcisin2014}, introduction of spatial parity violation (helicity) into the turbulent flow represents not only a more realistic physical scenario compared to the corresponding fully symmetric case but it additionally has the advantage of pronouncing different tensorial properties of the model under the study. Thus, based on the helical values of the corresponding Prandtl numbers a comparative analyses is performed in Ref.\,\cite{Jurcisin2014}. As a result, authors of Ref.\,\cite{Jurcisin2014} argue that structure of interactions exerts a more profound impact on diffusion-advection processes than the tensorial nature of the advected field itself. However, only three selected models are analyzed in Ref.\,\cite{Jurcisin2014} and strictly speaking the conclusions made by authors of Ref.\,\cite{Jurcisin2014} are merely hypotheses when extended beyond the range of the three studied models. The reason is that in Ref.\,\cite{Jurcisin2014} interactions could not be varied continuously. Nevertheless, kinematic MHD model and the aforementioned $A = 0$ model represent two special cases of the general $A$ model \cite{Jurcisin2014,Antonov2015,Antonov2015a,Jurcisin2016} with $A$ being a real parameter \cite{Arponen2009}. Thus, we may easily bring the kinematic MHD model and the $A = 0$ model of Ref.\,\cite{Jurcisin2014} onto a same footing by using the framework of the general $A$ model which allows direct description of a spectrum of different interactions by continuous variation of parameter $A$ (for details see Sec.\,\ref{sec2}). For $A = 1$ and $A = 0$ the two physically important cases of kinematic MHD and the $A = 0$ model of passively advected advection are recovered. Additionally, at $A = -1$ another important case of the so called linearized Navier-Stokes equations arises as a special case of the general $A$ model \cite{Arponen2009}. Thus, the general $A$ model represents a tool to unite several distinct but physically important cases into one single model. The advantage of such a generalization lies then in allowing for continuous variation of interaction structures which on the other hand greatly simplifies the analysis of influence of tensorial structures on diffusion-advection processes at least in the case of vector admixtures. 

A step towards such an analysis has already been undertaken in Ref.\,\cite{Jurcisin2016}, however only the case of fully symmetric turbulent environment has been considered and consequently only a limited insights have been gained into the problem. The assertions made by authors of Ref.\,\cite{Jurcisin2014} could therefore not be verified in Ref.\,\cite{Jurcisin2016}. It is therefore of high interest to analyze the general $A$ model with broken spatial parity and verify the hypothesis made in Ref.\,\cite{Jurcisin2014}. Moreover, the general $A$ model has also attracted a lot of attention recently from the point of view of their scaling properties, see for example Refs.\,\cite{Antonov2015,Antonov2015a,Antonov2003,Arponen2009}. But up to date only the case of fully symmetric turbulent environment has been analyzed. It is thus of high importance to include helical (violation of spatial parity) effects into the analysis of general $A$ model. For this purpose, it is the scope of the present paper to calculate for the first time the corresponding turbulent Prandtl number for the general $A$ model in fully developed turbulent environments with broken spatial parity. The resulting turbulent Prandtl number becomes then effectively a function of helical effects via the helical parameter $\rho$ (see Sec.\ref{sec2} for the definition) as well as function of the interaction parameter A. At this place, we also note that authors of Ref.\,\cite{Jurcisin2016} varied $A$ only in the range of $-1 \leq A \leq 1$ but according to Ref.\,\cite{Arponen2009} $A$ is in principle not bound to the interval $-1 \leq A \leq 1$. We therefore extend our analysis to the all possible values of $A$ but show later that constraints on $A$ arise artificially within the approach used in the present paper. 

To perform the investigations discussed above, we use the well established tools of field renormalization group (RG) technique as presented for example in Refs.\cite{VasilevBook,AdzhemyanBook,Zinn} which has widely been used in the field of fully developed turbulence without admixtures \cite{Adzhemyan1983,Adzhemyan2003,Adzhemyan2003a,Adzhemyan1988,Adzhemyan1996,Adzhemyan2005double,Adzhemyan2006double} as well as for advection diffusion processes of several admixtures including passive scalar admixture \cite{Adzhemyan1983a,Adzhemyan2005,Adzhemyan1998,Adzhemyan2001,Pagani2015,Novikov2003}, magnetic admixtures \cite{Adzhemyan1985,Adzhemyan1987} and also vector admixtures \cite{Jurcisin2014,Antonov2015,Antonov2015a,Jurcisin2016,Arponen2009,Adzhemyan2013,Arponen2010,Novikov2006}. Two loop techniques for calculation of the turbulent Prandtl number within the $A$ model used here are similar to those carried out in Ref.\,\cite{Adzhemyan2005}. The resulting helical values of turbulent Prandtl number are then analyzed to finally investigate the hypothesis raised by authors of Ref.\,\cite{Jurcisin2014}. In this respect, the context of the general $A$ model has also been used to further discuss the validity of two loop results on kinematic MHD obtained in Ref.\,\cite{Jurcisin2014}.

The paper is structured as follows. In Sec.\,\ref{sec2}, the $A$ model of passive advection of vector admixture is defined via stochastic differential equations. The emphasis is laid on the meaning of the parameter $A$ for the structure of interactions. In Sec.\,\ref{sec3}, field theoretic equivalent of stochastic differential equations of the $A$ model is introduced. The UV renormalization of the model is discussed in Sec.\,\ref{sec4} which is then concluded with the calculation of the IR stable fixed point of basic RG equations. Two loop calculation of the helical Prandtl number is presented in Sec.\,\ref{sec5} where also the helical dependence of the turbulent Prandtl number is discussed with special attention given to the influence of tensorial interaction structures on the diffusion advection processes in the $A$ model studied here. Obtained results are then briefly reviewed in Sec.\,\ref{sec6}.

\section{Model $A$ of passive vector advection with spatial parity violation}\label{sec2}

We consider a passive solenoidal vector field $\mathbf{b} \equiv \mathbf{b}(x)$ driven by a helical turbulent environment given by an incompressible velocity field $\mathbf{v} \equiv \mathbf{v}(x)$ where $x \equiv (t, \mathbf{x})$ with $t$ denoting the time variable and $\mathbf x$ the $d$ dimensional spatial position (later $d = 3$ strictly). Apparently, $\mathbf v$ and $\mathbf b$ are divergence free vector fields satisfying $\partial \, . \,  \mathbf{b} = \partial \, . \, \mathbf{v} = 0$. Additionally, within a general $A$ model of passive advection the following system of stochastic equations is required:
\begin{eqnarray}
\partial_t {\mathbf b}&=&\nu_0 u_0 \triangle {\mathbf b} -({\mathbf v}\cdot {\mathbf \partial}) {\mathbf b} + A({\mathbf b}\cdot {\mathbf
\partial}) {\mathbf v} -\partial P+ {\mathbf f^{b}}, \label{BB} 
\\
\partial_t {\mathbf v}&=&\nu_0 \triangle {\mathbf v} -({\mathbf v}\cdot {\mathbf \partial}) {\mathbf v} -\partial Q + {\mathbf f^v}, \label{vv}
\end{eqnarray}
where $\partial_t \equiv \partial / \partial_t$, $\partial_i \equiv \partial / \partial_{x_i}$, $\Delta \equiv \partial^2$ is the Laplace operator, $\nu_0$ is the bare viscosity coefficient, $u_0$ is the bare reciprocal Prandtl number, $P \equiv P (x)$ and $Q \equiv Q(x)$ represent the pressure fields while the stochastic terms $\mathbf f^v$, $\mathbf f^b$ and the  parameter $A$ are discussed later in this section. The subscript $0$ identifies unrenormalized quantities in what follows (see Sec.\,\ref{sec4} for more details).

Let us now briefly review the physical meaning of $A$ in Eq.\,(\ref{BB}). First, we note that Galilean symmetry requires $A$ only to be real with $A \in -1, 0, 1$ attracting most of the interest \cite{Jurcisin2016,Adzhemyan2013,Arponen2009,Arponen2010}. For $A = 1$ the kinematic MHD model is recovered, $A = 0$ leads to passive advection of a vector field in turbulent environments and finally $A = -1$ represents the model of linearized Navier-Stokes equations \cite{Arponen2009}. The parameter $A$ stands in front of the so called stretching term \cite{Adzhemyan2013} and due its continuous nature it represents a measure of specific interactions allowed by Galilean symmetry. Varying $A$ thus allows to investigate a variety of passively advected vector admixtures which only differ in their properties regarding interactions. According to \cite{Arponen2009}, parameter $A$ may take any real values but due to the special cases $A \in \{-1, 0, 1\}$ it is frequently only discussed in the smallest possible continuous interval encompassing all the three models, see for example Ref.\,\cite{Jurcisin2016}. Contrary, we extend the analysis to all physically allowed values of $A$, see Sec.\,\ref{sec5} for more details which allows a straightforward discussion of influence of interactions on advection diffusion processes.

The previously undefined stochastic terms $\mathbf f^v$ and $\mathbf f^b$ introduced in Eqs.\,(1) and (2) represent sources of fluctuations for $\mathbf v$ and $\mathbf b$. For energy injection of $\mathbf b$ we assume transverse Gaussian random noise $\mathbf{f^ b} = \mathbf{f^ b} (x)$ with zero mean via the following correlator:
\begin{equation}
D_{ij}^{b}(x;0)\equiv \langle f^b_i(x)f^b_j(0)\rangle=\delta(t)C_{ij}({\mathbf |{\mathbf x}|}/L), \label{CorelFb}
\end{equation}
where $L$ is an integral scale related to the corresponding stirring of $\mathbf b$ while $C_{ij}$ is required to be finite in the limit $L \rightarrow \infty$ and for $|x| \gg L $ it should rapidly decrease, but remains otherwise unspecified in what follows. Contrary, the transverse random force per unit mass $\mathbf{f^v} = \mathbf{f^v} (x)$ simulates the injection of kinetic energy into the turbulent system on large scales and must suit the description of real infrared (IR) energy pumping. To allow the later application of RG technique we shall assume a specific, power-like form of injection as usual for fully developed turbulence within the RG approach (for more details see Refs.\,\cite{VasilevBook,AdzhemyanBook,Adzhemyan1996}). Nevertheless, although a specific form is used universality of fully developed turbulence ensures that results obtained here may easily be extended to all fully developed turbulent flows. Additionally, it allows easy generalization to environments with broken spatial parity which is performed via tensorial properties of the correlator of $\mathbf{f^ v}$. For this purpose, we prescribe the following pair correlation function with Gaussian statistics:
\begin{eqnarray}
D_{ij}^v(x;0) &\equiv& \langle f^v_i(x) f^v_j(0) \rangle = \nonumber \\ &=& \delta(t)\int \frac{d^d {\mathbf k}}{(2\pi)^d} D_0 
k^{4-d-2\varepsilon} R_{ij}({\mathbf k}) e^{i {\mathbf k}\cdot {\mathbf x}}. \label{CorelFv}
\end{eqnarray}
Here, $d$ denotes the spatial dimension of the system, $\mathbf k$ is the wave number, $k$ denotes $|k|$, $D_0 ≡ g_0 \nu_0^3 > 0$ is the positive amplitude with $g_0$ being the coupling constant of the present model related to the characteristic ultraviolet (UV) momentum scale $\Lambda$ by the relation $g_0 \simeq \Lambda^{2\varepsilon}$. The term $R_{ij} (k)$ appearing in Eq.\,(\ref{CorelFv}) encodes the spatial parity violation of the underlying turbulent environment and its detailed structure is discussed separately in the text below. Finally, the parameter $\varepsilon$ is related to the exact form of energy injection at large scales and assumes value of $2$ for physically relevant infrared energy injection. However, as usual in the RG approach to the theory of critical behavior, we treat $\varepsilon$ formally as a small parameter throughout the whole RG calculations and only in the final step its physical value of $2$ is inserted  \cite{VasilevBook,Zinn}.

In Eq.\,(\ref{CorelFv}), we encounter typical momentum integrations which lead to two troublesome regions, namely the IR region of low momenta and UV region of high momenta as discussed in detail in Refs.\,\cite{VasilevBook,AdzhemyanBook}. Frequently, these troublesome integration regions are avoided by directly prescribing all relevant micro- and macroscopic properties of the flow. Here, we use the universality of fully developed turbulent flows to avoid unnecessary specifications. Thus, we only demand real IR energy injection of energy via Eq.\,(\ref{CorelFv}) and neglect the exact macroscopic structure of the flow by introducing a sharp IR cut-off $k \geq m$ for integrations over $\mathbf k$ with $L$ assumed to be much bigger than $1/m$. Using sharp cut-off, IR divergences like those in Eq.\,(\ref{CorelFv}) are avoided. As already done for Eq.\,(\ref{CorelFv}), the IR cut-off is understood implicitly in the whole paper and we shall stress out its presence only at the most crucial stages of the calculation. Contrary, UV divergences and their renormalization play central role in calculations presented here.

Finally, let us now turn our attention to the projector $R_{ij}$ in Eq.\,(\ref{CorelFv}) which controls all of the properties of the spatial parity violation in the present model. In the case of fully symmetric isotropic incompressible turbulent environments the projector $R_{ij} (k)$ assumes the usual form of the ordinary transverse projector
\begin{equation}
P_{ij}({\mathbf k})= \delta_{ij} - k_i k_j/k^2, \label{ProjectorP}
\end{equation}
see Ref.\,\cite{VasilevBook} for more details. In the case of helical flows, where spatial parity is violated, we specify Eq.\,(\ref{CorelFv}) in the form of a mixture of a tensor and a pseudotensor. Assuming isotropy of the flow we may divide the projector $R_{ij}$ in Eq.\,(\ref{CorelFv}) into two parts, i.e., $R_{ij} (k) = P_{ij} (k) + H_{ij} (k)$ where $H_{ij} (k)$ also respects the transversality of present fields. The ordinary non-helical transverse projector $P_{ij}$ is thus shifted by a helical contribution $H_{ij} (k)$ given as
\begin{equation}
H_{ij}({\mathbf k})=i \rho \, \varepsilon_{ijl} k_l/k. \label{ProjectorH}
\end{equation}
Here, $\epsilon_{ijl}$ is the Levi-Civita tensor of rank $3$ and the real parameter $\rho$ satisfies $|\rho| \in \left[ 0, 1 \right]$ due to the requirement of positive definiteness of the correlation function. Obviously, $\rho = 0$ corresponds to fully symmetric (non-helical) case whereas $\rho = 1$ means that parity is fully broken. The nonzero helical contribution leads to the presence of nonzero correlations $\langle \mathbf{v} . \it{rot} \, \mathbf{v} \rangle$ in the system.

We finally conclude the section by discussing the structure of interactions in Eqs.\,(\ref{BB}) and (\ref{vv}). Obviously, according to Eq.\,(ref{vv}) admixture field $\mathbf b$ does not disturb evolution of the velocity field $\mathbf v$. In other words, velocity field $\mathbf v$ is completely detached from the influence of admixtures as required by demanding passive advection. Of course, real problems usually involve at least some small amount of mutual interaction between the flow and its admixtures. However, even in the case of active admixtures there exist regimes which correspond to the passive advection problem as seen for example in the case of MHD problem with active magnetic admixture with its so-called kinetic regime controlled by the kinetic fixed point of the RG equations (see, e.g., Ref.\,\cite{Adzhemyan1985}). Such a situation corresponds to the passive advection obtained within the present model when $A = 1$ in Eqs.\,(\ref{BB}) and (\ref{vv}). The present picture of passive advection within the $A$ model represents thus a highly interesting physical scenario.

\section{Field theoretic formulation of the model \label{sec3}}

According to the Martin-Sigia-Rose theorem \cite{Martin}, the system of stochastic differential Eqs.\,(\ref{BB}) and (\ref{vv}) is equivalent to a field theoretic model of the double set of fields $\Phi = \{ v, b, v^{\prime}, b^{\prime} \}$ where unprimed fields correspond to the original fields of Eqs.\,(\ref{BB}) and (\ref{vv}) while primed fields are auxiliary response fields \cite{VasilevBook}. The field theoretic model is then defined via Dominicis-Janssen action functional
\begin{eqnarray}
S(\Phi) &=& \frac{1}{2} \int dt_1\,d^d{\mathbf x}_1\,dt_2\,d^d{\mathbf x}_2 \nonumber \\ && \hspace{-0.5cm} \Large[ v^{\prime}_i(x_1) D^v_{ij}(x_1;x_2) v^{\prime}_j(x_2) + b^{\prime}_i(x_1) D^b_{ij}(x_1;x_2) b^{\prime}_j(x_2) \Large] \nonumber   
\\
&+&  \int dt\,d^d{\mathbf x} \{ {\mathbf v^{\prime}} [-\partial_t +\nu_0 \triangle - ({\mathbf v} \cdot {\mathbf \partial}) ]{\mathbf v} \nonumber
\\ 
&& \hspace{-0.3cm} +\, {\mathbf b^{\prime}} [-\partial_t {\mathbf b} + \nu_0 u_0 \triangle {\mathbf b}  - ({\mathbf v}\cdot {\mathbf \partial}) {\mathbf b} + A ({\mathbf b}\cdot {\mathbf \partial}) {\mathbf v} ] \}, \nonumber
\\
\label{BareS} \,
\end{eqnarray}
where $x_l=(t_l,{\mathbf x}_l)$ with $l=1,2$, $D_{ij}^b$ and $D_{ij}^v$ are given in Eqs.\,(\ref{CorelFb}) and (\ref{CorelFv}), respectively, and required summations over dummy indices $i, j \in 1, 2, 3$ are implicitly assumed. Auxiliary fields and their original counterparts $\mathbf v$, $\mathbf b$ share the same tensor properties which means that all fields appearing in the present model are transverse. The pressure terms $\partial Q$ and $\partial P$ from Eqs.\, (\ref{BB}) and (\ref{vv}) respectively do not appear in action (\ref{BareS}) because transversality of auxiliary fields $\mathbf{v}^{\prime} (x)$ and $\mathbf{b}^{\prime} (x)$ allows to integrate these out of the action (\ref{BareS}) by using the method of partial integration.

\begin{figure}
\vspace{0.5cm}
\begin{center}\includegraphics[width=5cm]{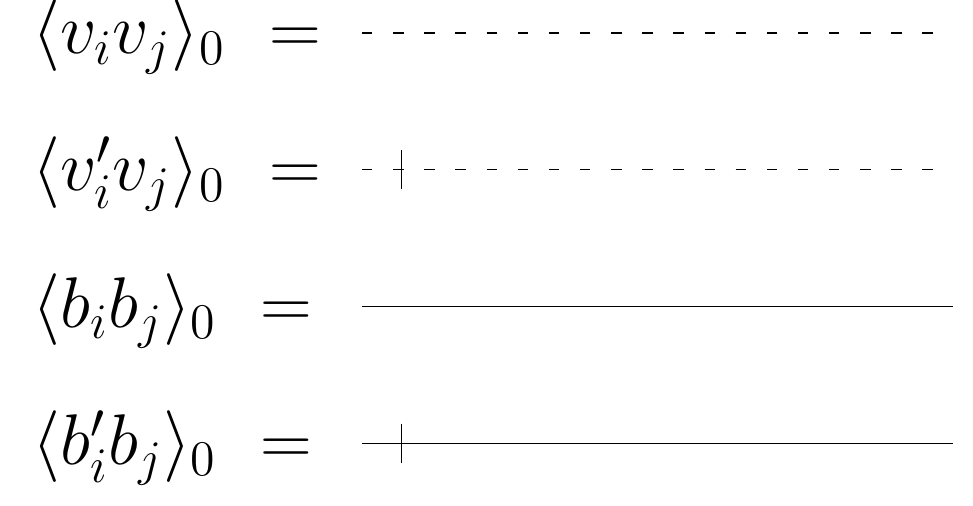}\end{center}
\caption{Graphical representation of the propagators of the model.
\label{fig1}}
\end{figure}

The field theoretic model of Eq.\,(\ref{BareS}) has a form analogous to the corresponding expression of Ref.\,\cite{Jurcisin2016} but includes via $D_{ij}^v$ the more general helical situation which was not considered by authors of Ref.\,\cite{Jurcisin2016}. In the frequency-momentum representation the following set of bare propagators is obtained:
\begin{eqnarray}
\langle b_i^{\prime} b_j \rangle_{0} = \langle b_i b_j^{\prime} \rangle_{0}^* & = & \frac{P_{ij}({\mathbf k})}{i\omega+\nu_{0} u_0 k^{2}},\label{Propagator_Bb} 
\\
\langle v_i^{\prime} v_j \rangle_{0} = \langle v_i v_j^{\prime} \rangle_{0}^* & = & \frac{P_{ij}({\mathbf k})}{i\omega+\nu_{0} k^{2}},\label{Propagator_Vv} 
\\
\langle b_i b_j\rangle_{0} & = & \frac{C_{ij}({\mathbf k})}{|-i\omega+\nu_{0} u_0 k^{2}|^2},\label{Propagator_bb} 
\\
\langle v_i v_j\rangle_{0} & = & \frac{g_0 \nu_0^3 k^{4-d-2\varepsilon}R_{ij}({\mathbf k})}{|-i\omega+\nu_{0} k^{2}|^2},\label{Propagator_vv}
\end{eqnarray}
with helical effects already appearing in the propagator (\ref{Propagator_vv}). Function $C_{ij}(k)$ is the Fourier transform of the function $C_{ij} (r/L)$ which appears in Eq.\,(\ref{CorelFb}), but remains arbitrary in the calculations that follow. Propagators are represented as usual by dashed and full lines, where dashed lines involve velocity type of fields and full lines represent vector admixture type fields. Auxiliary fields are denoted using a slash in the corresponding propagators as shown in Fig.\,\ref{fig1} \cite{VasilevBook}.

Field theoretic formulation of the $A$ model contains also two different triple interaction vertices, namely $b_i^{\prime}(-v_{j}\partial_{j}b_i+ A b_j\partial_j v_i)=b_i^{\prime}v_{j}V_{ijl} b_l$ and $-v_i^{\prime}v_{j}\partial_{j} v_i=v_i^{\prime}v_{j}W_{ijl} v_l/2$. In the momentum-frequency representation, $V_{ijl}=i(k_j\delta_{il}-A k_l \delta_{ij})$  while $W_{ijl}=i(k_l \delta_{ij}+k_j\delta_{il})$. In both cases, momentum $\mathbf k$ is flowing into the vertices via the auxiliary fields ${\mathbf b^{\prime}}$ and ${\mathbf v^{\prime}}$, respectively. In the end, let us also briefly remind that formulation of the stochastic problem given by Eqs.\,(\ref{BB})-(\ref{vv}) through the field theoretic model with the action functional (\ref{BareS}) allows one to use the well-defined field theoretic means, e.g., the RG technique, to analyze the problem \cite{VasilevBook,Collins}. 

\begin{figure} [t]
\vspace{0.5cm}
\begin{center}\includegraphics[width=9.cm]{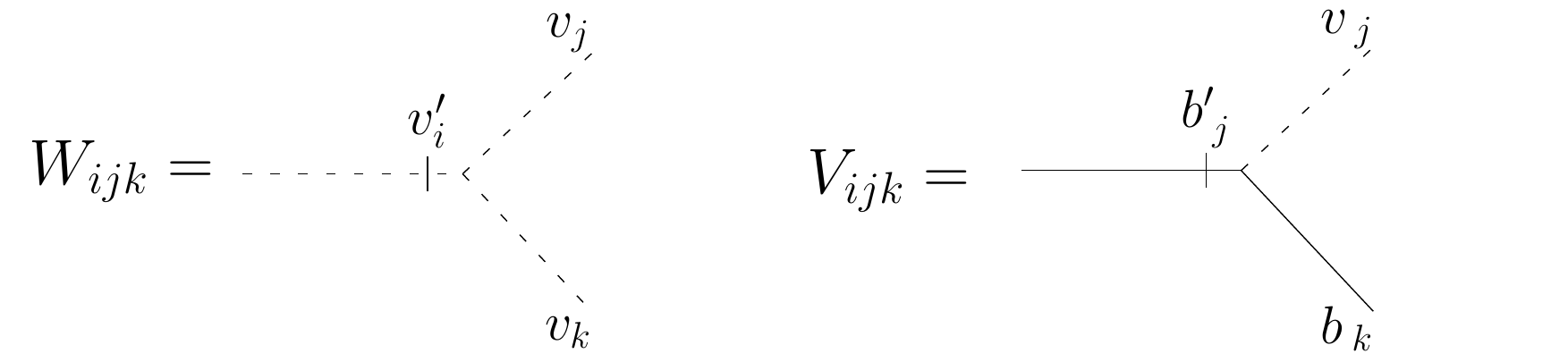}\end{center}
\caption{Two interaction vertices of the $A$ model. The $W_{ijk}$ type of vertex involves only velocity type of fields $\mathbf{v}$ and $\mathbf{v^{\prime}}$ with $W_{ijl}=i(k_l \delta_{ij}+k_j\delta_{il})$. The second interaction vertex $V_{ijk}$ is the only basic diagrammatic object of the corresponding Feynman rules for present model which contains $A$ dependent contribution in the form of $V_{ijl}=i(k_j\delta_{il}-A k_l \delta_{ij})$. \label{fig2}}
\end{figure}

\section{Renormalization group analysis} \label{sec4}

The RG analysis performed here requires to determine all relevant UV divergences in the present model. Therefore, we employ the analysis of canonical dimensions which allows to identify all objects (graphs) containing the so called superficial UV divergences as they turn out to be the only relevant divergences left for the subsequent RG analysis in the present paper. For details, see Refs.\,\cite{VasilevBook,AdzhemyanBook,Zinn}. 

Since the present $A$ model belongs to the class of the so called two scale models \cite{VasilevBook,AdzhemyanBook,Adzhemyan1996}, an arbitrary quantity $Q$ has a canonical dimension $d_Q = d^k_Q + d^{\omega}_Q$, where $d_Q^k$ corresponds to the canonical dimension of $Q$ connected with the momentum scale and $d^{\omega}_Q$ corresponds to the frequency scale. Our general helical model differs from the simple model studied in Ref.\,\cite{Jurcisin2016} by inclusion of $\rho$ dependent terms which encode helical effects of turbulent environments with broken parity. Therefore, non-helical results of Ref.\,\cite{Jurcisin2016} have to be carefully reexamined for the present model. Nevertheless, in the limit of $\rho \rightarrow 0$, the general helical $A$ model has to give the same results as its non-helical counterpart of Ref.\,\cite{Jurcisin2016}. A straightforward calculation of canonical dimensions in the present model shows that $d^k_{\rho} = d^{\omega}_{\rho} = 0$ while all the other remaining quantities posses canonical dimensions as in Ref.\,\cite{Jurcisin2016}. 

In conclusion, analysis of canonical dimensions shows that the helical $A$ model posses dimensionless coupling constant $g_0$ at $\varepsilon = 0$. The present model is thus logarithmic at $\varepsilon = 0$ which means that in the framework of minimal subtraction scheme, as used in what follows, all possible UV divergences are of the form of poles in $\varepsilon$ \cite{Zinn,Collins}. Then, using the general expression for the total canonical dimension of an arbitrary 1-irreducible Green's function $\langle \Phi \ldots \Phi \rangle_{1-ir}$, which plays the role of the formal index of the UV divergence, together with the symmetry properties of the model, one finds that for physical dimension $d=3$, the superficial UV divergences are present only in the 1-irreducible Green's functions $\langle v_i^{\prime} v_j \rangle_{1-ir}$ and$\langle b_i^{\prime} b_j \rangle_{1-ir}$. Thus, all divergences can be removed by counterterms of the forms $\mathbf{v}^{\prime} \Delta \mathbf{v}$ or $\mathbf{v}^{\prime} \Delta \mathbf{v}$ which leads to multiplicative renormalization of the parameters $g_0$, $u_0$, and $\nu_0$ via renormalization constants $Z_i = Z_i (g, u; d, \rho; \varepsilon)$ as
\begin{equation}
\nu_{0}=\nu Z_{\nu}, \quad g_{0}=g\mu^{2\varepsilon}Z_{g}, \quad u_{0}=u Z_{u},\label{RenormVGU}
\end{equation}
where the dimensionless parameters $g,u$, and $\nu$ are the renormalized counterparts of the corresponding bare ones and $\mu$ is the renormalization mass required for dimensional regularization as used in the present paper. Quantities $Z_i = Z_i (g, u; d, \rho; \varepsilon)$ contain poles in $\varepsilon$.

However, there exist one additional problem when passing from the non-helical to the general parity broken $A$ model. Strictly speaking, the above conclusions are completely true only in the non-helical case. In the general case ($0 < |\rho| \leq 1$), linear divergences in the form of $\mathbf{b^{\prime}} . rot \, \mathbf{b}$ appear in the 1-irreducible Green’s function $\langle b^{\prime}_i b_j \rangle_{1-ir}$, see Ref.\,\cite{Adzhemyan1987} for more details. Removing them multiplicatively, corresponding linear terms would have to be introduced into the action functional. On the other hand, such new terms would lead to the instability causing the exponential growth in time of the response function $\langle b^{\prime}_i b_j \rangle$. A correct treatment inherently requires a genuine interplay between the underlying helical velocity field and its admixtures which is beyond the scope of passive advection $A$ model. Therefore, we shall leave the problem of the linear divergences untouched in the present paper and concentrate only on the problem of the existence and stability of the IR scaling regime, which can be studied without considering the linear divergences as already done for similar problem for example in Ref.\,\cite{Jurcisin2014}. However, we stress out that the full problem can only be solved when $A$ model with active admixtures is considered which should be the next logical in continuing the present analysis to more complicated systems. 

Bearing the problem of linear $\rho$ divergences in mind, we continue the RG analysis by writing the renormalized action functional as 
\begin{eqnarray}
S(\Phi) &=& \frac{1}{2} \int dt_1\,d^d{\mathbf x}_1\,dt_2\,d^d{\mathbf x}_2 \nonumber \\ && \hspace{-0.5cm} \Large[ v^{\prime}_i(x_1) D^v_{ij}(x_1;x_2) v^{\prime}_j(x_2) + b^{\prime}_i(x_1) D^b_{ij}(x_1;x_2) b^{\prime}_j(x_2) \Large] \nonumber   
\\
&+&  \int dt\,d^d{\mathbf x} \{ {\mathbf v^{\prime}} [-\partial_t +\nu Z_1 \triangle - ({\mathbf v} \cdot {\mathbf \partial}) ]{\mathbf v} \nonumber \\ && \hspace{-0.3cm} +\, {\mathbf b^{\prime}} [-\partial_t {\mathbf b} + \nu u Z_2 \triangle {\mathbf b}  - ({\mathbf v}\cdot {\mathbf \partial}) {\mathbf b} + A  ({\mathbf b}\cdot {\mathbf \partial}) {\mathbf v} ] \}, \nonumber 
\\
\label{BareSr}
\end{eqnarray}
with $Z_1$ and $Z_2$ being the renormalization constants connected with the previously defined renormalization constants $Z_i = Z_i (g, u; d, \rho; \varepsilon)$ with $i \in \nu, g, \mu$ via equations
\begin{equation}
Z_{\nu}=Z_{1},\quad  Z_{g}=Z_{1}^{-3},\quad Z_{u}=Z_2
Z_{1}^{-1}.\label{Renorm12}
\end{equation}
Each of the renormalization constants $Z_ 1$ and $Z_2$ corresponds to a different class of Feynman diagrams (as discussed below) but they share an analogous structure within the MS scheme: the $n$-th order of perturbation theory corresponds to $n$-th power of $g$ with the corresponding expansion coefficient containing a pole in $\varepsilon$ of multiplicity $n$ and less, i. e.:
\begin{eqnarray}
Z_{1}(g;d,\rho;\varepsilon) &=& 1+\sum_{n=1}^{\infty} g^n \sum_{j=1}^n \frac{z^{(1)}_{nj}(d,\rho)}{\varepsilon^j}, \label{Z1expansion} 
\\
Z_{2}(g,u;d,\rho;\varepsilon) &=& 1+\sum_{n=1}^{\infty} g^n \sum_{j=1}^n \frac{z^{(2)}_{nj}(u,d,\rho)}{\varepsilon^j}, \label{Z2expansion}
\end{eqnarray}
where we defined $\varepsilon$ independent terms $z^{(1)}_{nj}(d,\rho)$ and $z^{(2)}_{nj}(u,d,\rho)$ and explicitly divided them by corresponding poles over $\varepsilon$. Using the last expressions with renormalized variables inserted leads to divergence free 1-irreducible Green's functions $\langle  v_i^{\prime} v_j \rangle_{1-ir}$ and $\langle b_i^{\prime} b_j \rangle_{1-ir}$. Moreover, 1-irreducible Green's functions $\langle v_i^{\prime} v_j \rangle_{1-ir}$ and $\langle b_i^{\prime} b_j \rangle_{1-ir}$  are associated with the corresponding self-energy operators $\Sigma^{v^{\prime}v}$ and $\Sigma^{b^{\prime} b}$ by the Dyson equations which in frequency-momentum representation read
\begin{eqnarray}
\langle v_i^{\prime}v_j \rangle_{1-ir}&=&[ \, \, \, i\omega \, \, \, -\nu_0 p^2 + \Sigma^{v^{\prime}v}(\omega, p)] P_{ij}({\mathbf p}),\label{Dyson1} 
\\
\langle b_i^{\prime}b_j \rangle_{1-ir}&=&[i\omega-\nu_0 u_0 p^2+ \Sigma^{b^{\prime}b}(\omega, p)]P_{ij}({\mathbf p}).\label{Dyson2}
\end{eqnarray}
Thus, substitution of $e_{0}= e \mu^{d_e} Z_{e}$ for $e=\{g,u,\nu \}$ is required to lead to UV convergent Eqs.\,(\ref{Dyson1}) and (\ref{Dyson2}) which in turn determine the renormalization constants $Z_1$ and $Z_2$ up to an UV finite contribution. However, by choosing the minimal subtraction (MS) scheme in what follows we require all renormalization constants have the form of 1 + \textit{poles in $\varepsilon$}. In the end, one gets explicit expressions for coefficients $z_{nj}^{(i)}$, $i = 1, 2$ in Eqs.\,(\ref{Z1expansion}) and (\ref{Z2expansion}) in the corresponding order of the perturbation theory. As explained earlier, only logarithmic divergences are considered within the general $A$ model of passive advection and possible linear divergences in $\rho$ remain untreated.

The aim of the present paper consists of deriving two-loop perturbative results for the $A$ model with helical effects included via proper definition of Eq.\,(\ref{CorelFv}). Since in the limit $\rho \rightarrow 0$ the less general non-helical model of Ref.\,\cite{Jurcisin2016} is recovered, all non-helical results of Ref.\,\cite{Jurcisin2016} have to be reproduced here. Moreover, all quantities depending exclusively on velocity field $\mathbf v$ follow only from stochastic Navier-Stokes equation (\ref{vv}) and the correlator (\ref{CorelFv}). In Refs.\,\cite{Jurcisin2014,Adzhemyan1988}, exactly the same conditions have been imposed on velocity type of fields $\mathbf{v}$ and $\mathbf{v^{\prime}}$ in two loop calculations of the given model. Consequently, the corresponding quantities depending exclusively on velocity type of fields in the present paper model have to equal those obtained in Refs.\,\cite{Jurcisin2014,Adzhemyan1988}. Taking together, $Z_1$ in the present model must be the same as in Ref.\,\cite{Adzhemyan1988} while non-helical values of $Z_2$ in the generalized helical $A$ model must reproduce results of Ref.\,\cite{Jurcisin2016}. Thus, before generalizing the approach of Refs.\,\cite{Jurcisin2016,Adzhemyan2005} to the more general $A$ model with helical contributions, we review results of Refs.\,\cite{Jurcisin2014,Jurcisin2016,Adzhemyan2005} which are relevant for the present paper.

Let us start with coefficients related to the underlying turbulent environment given by $\mathbf v$ which comprise the renormalization coefficient $Z_1$. As stated above, the present model and the model under the study in Refs.\,\cite{Jurcisin2014,Adzhemyan1988} have the same renormalization constant $Z_1$. Its one-loop expansion coefficient $z_{11}^{(1)}$ therefore reads
\begin{equation}
z^{(1)}_{11}=-\frac{S_d}{(2\pi)^d}\frac{(d-1)}{8(d+2)}, \label{z1_11}
\end{equation}
where $S_d$ is the surface area of the $d$-dimensional unit sphere defined as $S_d \equiv 2 \pi^{d/2} / \Gamma(d/2)$ with $\Gamma(x)$ being the standard Euler's Gamma function. Thus, no helical contributions at one-loop level emerge for quantities involving only velocity type of fields $\mathbf{v}$ and $\mathbf{v^{\prime}}$. The two loop order coefficient $z^{(1)}_{22}$ is in Ref.\,\cite{Adzhemyan1988} shown to satisfy
\begin{equation}
z^{(1)}_{22}=-\left(z^{(1)}_{11}\right)^2, \label{z1_22}
\end{equation}
Consequently, $z^{(1)}_{22}$ is actually also $\rho$ independent. Thus, only the remaining coefficient $z^{(1)}_{21}$ contains helical contributions to $Z_1$ . Nevertheless, the corresponding expression from Ref.\,\cite{Adzhemyan2005} is rather huge and we shall not reprint it here. 

\begin{figure}
\vspace{0.5cm}
\begin{center}\includegraphics[width=8.5cm]{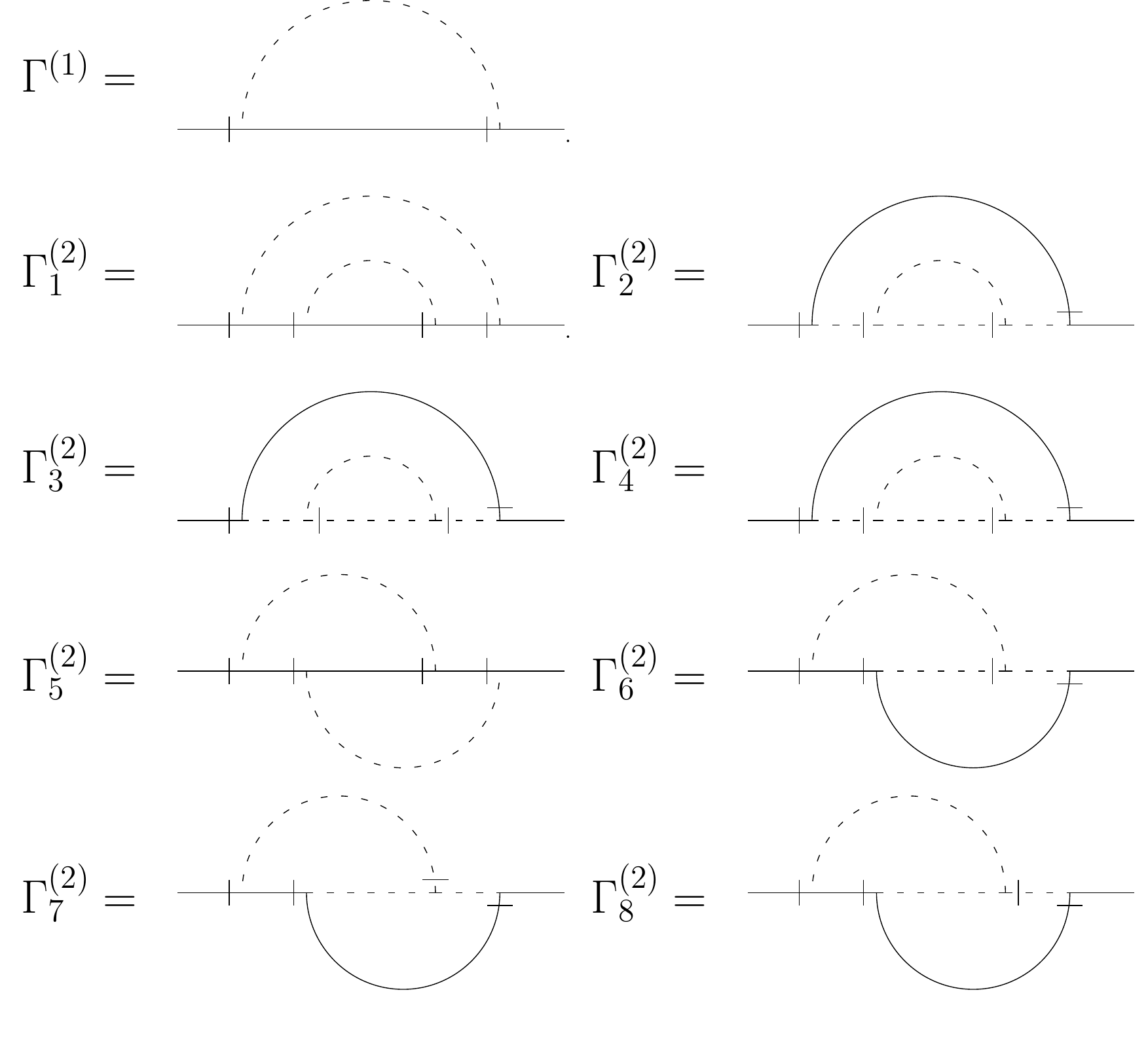}\end{center}
\caption{One-loop and two-loop diagrams that contribute to the self-energy operator $\Sigma^{b^{\prime}b} (\omega, p)$ in Eq.\, (\ref{Dyson2}).
\label{fig3}}
\end{figure}

Let us now reexamine the calculations of $Z_2$ done by authors of Ref.\,\cite{Jurcisin2016} with special attention given to the extension of the procedure to the more general helical $A$ model of passive advection as considered here. For this purpose we shall analyze the structure of the self-energy operator $\Sigma^{b^{\prime} b}$ in the Dyson equation (\ref{Dyson2}). In the two loop order, $\Sigma^{b^{\prime} b}$ equals the sum of singular parts of nine one-irreducible Feynman diagrams as shown in Fig.\,{\ref{fig3}}. Using the notation of Ref.\,\cite{Jurcisin2016} for the sake of easier comparison, we write down the two-loop approximation of $\Sigma^{b^{\prime} b}$ as
\begin{equation}
\Sigma^{b^{\prime} b} = \Gamma^{(1)} + \Gamma^{(2)} = \Gamma^{(1)} + \sum\limits_{l=1}^{8} s_l \Gamma^{(2)}_l \label{Dyson2structure}
\end{equation}
where $\Gamma(1)$ represents the single one-loop diagram shown in Fig.\,\ref{fig3} and $\Gamma^{(2)}$ represents the sum of eight two-loop diagrams shown in Fig.\,\ref{fig3}. Terms $s_l$, $l = 1, \ldots, 8$ denote the corresponding symmetry factors which equal $1$ for all diagrams except of the fourth with $s_4 = 1/2$. 

The single one loop diagram of Fig.\,\ref{fig3} apparently does not include the propagator $\langle v_i v_j \rangle_0$  which is the only diagrammatic object that contains helical contributions. The corresponding coefficient $z_{11}^{(2)}$ that follows from the $\Gamma^{(1)}$ contribution is thus actually also $\rho$ independent. Since all non-helical quantities in the present helical $A$ model must reproduce the corresponding values of Ref.\,\cite{Jurcisin2016} the following $z_{11}^{(2)}$ expansion coefficient must be obtained (as verified also by direct calculation):
\begin{eqnarray}
z^{(2)}_{11} &=& -\frac{S_d}{(2\pi)^d} \nonumber 
\\
&\times& 
\frac{ (d^2-3) (u+1) + A \left[ d + u (d-2) \right] + A^2 (1 + 3u) }{4d(d+2)u(u+1)^2 } \nonumber 
\\ \label{z2_11}
\end{eqnarray}
which in the case of $A = -1$, a special case focused later on in the paper, simplifies to 
\begin{equation}
z^{(2)}_{11}=-\frac{S_d}{(2\pi)^d}\frac{(2u-1) +d(d-1)(u+1) }{4 d (d+2) u (u+1)^2}.
\label{z2_11a}
\end{equation}
Let us now analyze the contributions to $\Gamma^{(2)}$ which determine $z^{(2)}_{22} (d, \rho)$ and $z^{(2)}_{21} (d, \rho)$. As already stated, there are eight two loop diagrams contributing to $\Gamma^{(2)}$. After a quick inspection we notice that each of the diagrams contains two $ \langle v_i v_j \rangle_0$ propagators which are linearly dependent on helicity parameter $\rho$. Thus, all two loop diagrams can depend only quadratically on $\rho$ (linear dependencies are not relevant for present calculations and are dropped systematically). Thus, using notation equivalent to that of Ref.\,\cite{Jurcisin2016} we may write the divergent part of $\Gamma^{(2)}$ in the following form:
\begin{eqnarray}
\Gamma^{(2)}&=&\frac{g^2\nu\,p^2\,S_d}{16(2\pi)^{2d}} \left(\frac{\mu}{m}\right)^{4\varepsilon} \frac{1}{\varepsilon}
\nonumber 
\\ 
&\times& \Bigg\{ \frac{S_d}{\varepsilon} C^{\rho}  + B^{(0)} + \rho^2 \delta_{3d} B^{(\rho)}  \Bigg\}, \label{Gamma2structure}
\end{eqnarray}
where $C^{\rho}$, $B^{(0)}$ and $B^{(\rho)}$ are for now on undetermined. We note that $d, g, p, \mu, u, m$ dependent factors in Eq.\,(\ref{Gamma2structure}) could principally by absorbed into $C^{\rho}$, $B^{(0)}$ and $B^{(\rho)}$, but in order to comply with the notation of Ref.\,\cite{Jurcisin2016} the specific form of Eq.\,(\ref{Gamma2structure}) is used. Since by definition, $B^{(0)}$ encodes the non-helical contributions of the corresponding diagrams we notice that it must yield the same result as obtained in Ref.\,\cite{Jurcisin2016}. However, $B^{(0)}$ was not explicitly introduced in Ref.\,\cite{Jurcisin2016} but it may easily be expressed via following equation:
\begin{eqnarray}
B^{(0)} &=& S_{d-1} \int_0^1 d x\,\, (1-x^2)^{(d-1)/2} \,\, B, \label{Bzero}
\end{eqnarray}
where variable $x$ denotes the cosine of the angle between two independent loop momenta $\mathbf k$ and $\mathbf q$ of the two-loop diagrams, i.e., $x ={\mathbf k}.{\mathbf q}/|k|/|q|$ and $B$ is a function explicitly introduced by authors of Ref.\,\cite{Jurcisin2016}. Nevertheless, $B$ is a complicated function of $u$ and $A$ as shown in Appendix of Ref.\,\cite{Jurcisin2016} and shall not be reprinted here. We merely notice that within the scope of the present calculations we have determined $B^{(0)}$ directly by methods discussed later in connection with helical contributions in the present model. We state in advance that the special non-helical values of Ref.\,\cite{Jurcisin2016}, expressed via Eq.\,(\ref{Bzero}) above, have been confirmed to hold within the present general helical $A$ model. On the other hand, the expression $C^{\rho}$ is directly related to the second order pole coefficient of $Z_2$, namely to $z_{11}^{(2)} (d, \rho)$. Although, we denoted this contribution with superscript $\rho$, in reality it must be independent of helical contributions when divergences linear in $\rho$ are left untouched as done in the present paper. The reason for vanishing of the possible $\rho$ dependence lies in the one-loop order of the present generalized $A$ model which is completely free of any helical effects. Consequently, second order $\varepsilon$ pole contributions to $\Gamma^{(2)}$ have to remain also $\rho$ independent. Particularly, it means that superscript $\rho$ in $C^{\rho}$ may be dropped, i. e., $C^{\rho} \equiv C$. Because $\rho$ dependencies are not present in $C^{\rho} \equiv C$ it must equal to Eq.(32) of Ref.\,\cite{Jurcisin2016} yielding thus the corresponding $z_{22}^{(2)} (d, \rho)$ actually also $\rho$ independent:
\begin{equation}
z^{(2)}_{22} (d,\rho) = z^{(2)}_{22} (d) = -\frac{S_d^2}{(2\pi)^{2d}} \frac{C}{16u}. \label{z2_22}
\end{equation}
The coefficient $C^{\rho} \equiv C$ may be calculated directly. At this place, we only review its form and postpone the details of calculation for later on. In accordance to Ref.\,\cite{Jurcisin2016} and to calculations performed within the scope of the present article, $C^{\rho} \equiv C$ is a polynomial of fourth order in $A$ while corresponding coefficients are complicated rational functions of $d$ and $u$ and shall not be reprinted here, for details see Eq.\,(32) in Ref.\,\cite{Jurcisin2016}. 

Taking together, in Eqs.\,(\ref{z1_11})-(\ref{z2_22}) we have briefly discussed results common among the present model and models of Refs.\,\cite{Jurcisin2014,Jurcisin2016}. Passing to our generalized helical $A$ model requires now an explicit calculation of helical contributions to $\Gamma^{(2)}$ . Now, we stress out that although $B^{(\rho)}$ is calculated with explicit $d$ dependence, the helical contributions make only sense for $d = 3$ as already stated several times and made explicit by insertion of Kronecker delta $\delta_{d3}$ into the Eq.\,(\ref{Gamma2structure}). 

However, before going further, let us now explain the general character of $A$ dependencies in expressions $C^{\rho} \equiv C$, $B^{(0)}$ and $B^{(\rho)}$ without considering the details of the corresponding calculations. According to Fig.\,\ref{fig3} and Eqs.(\ref{Dyson2structure}) and (\ref{Gamma2structure}), all of the discussed expressions are connected with diagrams $\Gamma^{(1)}$ or $\Gamma^{(2)}_l$ with $l = 1, \ldots 8$. Noting now that parameter $A$ appears only in the $V_{ijl}$ type vertex as a linear function we may gain direct insights into the structure of $A$ dependencies of given diagrams. To this end, imagine now a diagram with only two vertices of $V_{ijl}$ type. Since each of the vertices contains only a linear function of $A$ when necessary summations on dummy field indices are performed we get an overall dependence which may include the most a quadratical term in $A$ as a result of two linear terms in $A$ being multiplied together. In other words, the resulting diagram may therefore be only a polynomial in $A$ of order $2$ the most. The same reasoning extends also to the case when four $V_{ijl}$ type vertices appear simultaneously in given diagram. Here, the resulting polynomial must be of order four in $A$. Of coarse, since $V_{ijl}$ type vertices are of tensorial nature, summation over field indices in given diagram may lower the actual order of the polynomials in $A$ while some polynomial coefficients may also vanish completely. However, under any circumstances higher powers of $A$ may not emerge in the graphs. Using now the previous conclusions, diagrams $\Gamma_l$ with $l = 1, \ldots 8$ contain two or four $V_{ijl}$ type vertices and their sum $\Gamma^{(2)}$ must consequently be a polynomial in $A$ of the order $4$ the most. Subsequently, since $C^{\rho} \equiv C$ is proportional to the second order pole in $\varepsilon$ of $\Gamma^{(2)}$ it must also be a polynomial of the order $4$ the most. Parameters $B^{(0)}$ and $B^{(\rho)}$ are proportional to the corresponding parts of $\Gamma^{(2)}$ and must therefore also be polynomials in $A$ with order $4$ the most. 

Although previous discussions determine the structure of the diagrams, only direct calculation may give us the needed coefficients of the resulting polynomials in $A$. Thus, we have to perform the calculation of the coefficients $z_{21}^{(2)} (u,d, \rho)$ and $z_{22}^{(2)} (d, \rho)$ directly. As already seen, $z_{21}^{(2)} (u,d, \rho)$ has to comply with Eq.\,(\ref{z2_21nohelicity}) in the limit $\rho \rightarrow 0$. On the other hand, since all helical properties of the generalized helical $A$ model are encoded by the term $B^{(\rho)}$ and linear $\rho$ divergences are left out in the passive advection within the $A$ model we already note that $z_{21}^{(2)} (u,d, \rho)$ contains a quadratic term in $\rho$ as the only $\rho$ dependent part. However, to correctly determine the exact term proportional to $\rho^2$ we are required to calculate $B^{(\rho)}$. For this purpose, we use the Dyson equation (\ref{Dyson2}), the relation (\ref{Dyson2structure}), and the structure of $\Gamma^{(2)}$ as given by Eq.\,(\ref{Gamma2structure}). In the end, $z_{21}^{(2)}(u,d,\rho)$ is found as (once again notation of Ref.\,\cite{Jurcisin2016} is used)
\begin{equation}
z_{21}^{(2)} (u,d, \rho) = \frac{S_d S_{d-1}}{16 u (2\pi)^{2d}} \left( B^{(0)} + \rho^2 \delta_{d3} B^{(\rho)} \right),
\label{z2_21helical}
\end{equation}
where $B^{(0)}$ and $B^{(\rho)}$ are defined via Eqs.\,(\ref{Gamma2structure}) and (\ref{Bzero}), respectively. According to Eq.\,(\ref{z2_21helical}), $B^{(\rho)}$ is given by eight two loop diagrams of Fig.\,\ref{fig3} which have a graphical representation equal to that of Refs.\,\cite{Jurcisin2014,Jurcisin2016,Adzhemyan2005} but are inherently different because of helical effects included via the propagator $\langle v_i v_j \rangle_0$. In close analogy to Eq.\,(\ref{Bzero}) we write $B^{(\rho)}$ as
\begin{eqnarray}
B^{(\rho)} &=& S_{d-1} \int_0^1 d x\,\, (1-x^2)^{(d-1)/2} \,\, \sum_{l=1}^8 s_l B^{(\rho)}_{l}, \label{Brho}
\end{eqnarray}
and define thus $B^{(\rho)}_{l}$ to be helical contributions from the corresponding parts of $\Gamma_|^{(2)}$ diagrams. Thus, as already discussed, when the limit $\rho \rightarrow 0$ is imposed on Eq.\,(\ref{z2_21helical}) the resulting value gives the $B^{(0)}$ coefficient which then in turn complies with its corresponding counterpart of Ref.\,\cite{Jurcisin2016}. On the other hand, for $\rho \neq 0$ the eight two-loop graphs contain nonzero terms which then via $B^{(\rho)}$ encode all of the helical effects investigated here. In other words, result of Ref.\,\cite{Jurcisin2016} are only a special case of the present calculations when appropriate limits are taken while for $0 < |\rho| \leq 1$ the corresponding expressions are completely unknown and require to be calculated here. For this purpose, for diagrams $\Gamma^{(2)}_l$ with $l = 2, \ldots 8$, we utilize the derivative technique outlined in Ref.\,\cite{Adzhemyan2005} whose prerequisites are fulfilled for selected diagrams with $l = 2, \ldots 8$. However, in the case of diagram $\Gamma^{(2)}_1$, only its non-helical value, a special case of the model considered here, can by evaluated using the derivative technique of Ref.\,\cite{Adzhemyan2005}. Therefore, the well established techniques outlined for example in Ref.\,\cite{VasilevBook} are used for the remaining graph $\Gamma^{(2)}_1$. Nevertheless, calculations for all graphs are quite straightforward, however they result in complicated lengthy expressions and we present them in the Appendix of the present paper. 

In the end, we have to reexamine the influence of helicity on the properties of the IR scaling regime and its stability. First of all, since fields $\mathbf v$, $\mathbf v^{\prime}$, $\mathbf b$, and $\mathbf b^{\prime}$ are not renormalized the following simple relation is satisfied: 

\begin{figure}
\vspace{0.5cm}
\begin{center}\includegraphics[width=8cm]{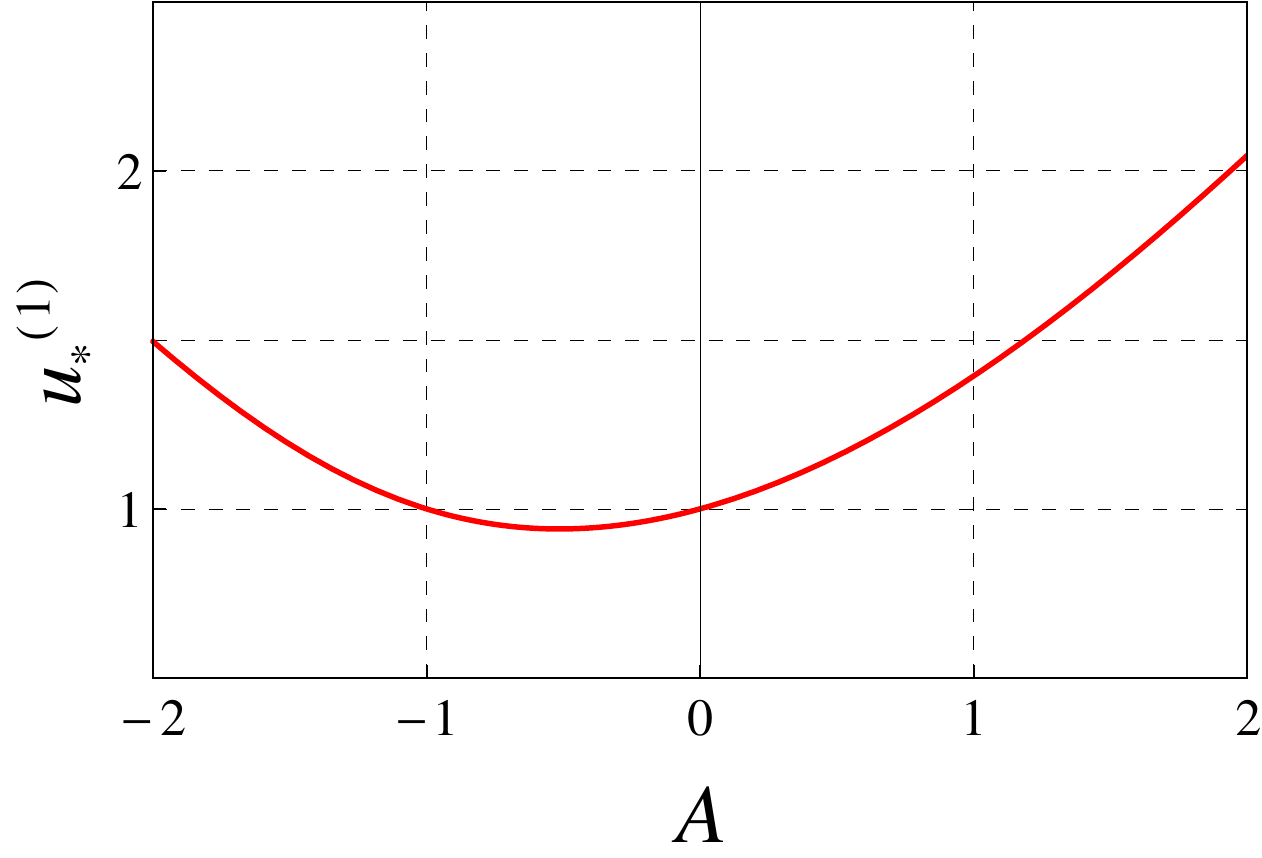}\end{center}
\caption{ (Color online) Dependence of one loop inverse turbulent Prandtl number $u^{(1)}_{*}$ on parameter $A$ in region $-2 \leq A \leq 2$. Note that for $A = -1$ one obtains $u_{*}^{(1)} = 1$. Apparently, one loop values of $u_{*}^{(1)}$ are always positive ($u_{*}^{(1)} \rightarrow \infty$ for $A \rightarrow \pm \infty$) and therefore physical for all arbitrary real $A$. 
\label{fig4}}
\end{figure}

\begin{equation}
W^R(g,u,\nu,\mu,\cdots)=W(g_0,u_0,\nu_0,\cdots), \label{RenormW}
\end{equation}
It states that renormalized connected correlation functions $W^R=\langle \Phi \dots \Phi \rangle^R$ differ from their unrenormalized counterparts $W=\langle \Phi \dots \Phi \rangle$ only by the choice of variables (renormalized or unrenormalized) and in the corresponding perturbation expansion (in $g$ or $g_0$), where dots stand for other arguments which are untouched by renormalization, e.g., the helicity parameter or coordinates \cite{VasilevBook,AdzhemyanBook,Collins}. This however means that unrenormalized correlation functions are independent of the scale-setting parameter $\mu$ of dimensional regularization. Thus, applying the differential operator $\mu \partial_{\mu}$ at fixed unrenormalized parameters on both sides of Eq.\,(\ref{RenormW}) gives the basic differential RG equation of the following form \cite{VasilevBook,AdzhemyanBook}:
\begin{equation}
[\mu \partial_{\mu} + \beta_g\partial_g + \beta_u\partial_u-\gamma_{\nu} \nu
\partial_{\nu}] W^R(g,u,\nu,\mu,\cdots)=0, \label{BasicRG}
\end{equation}
where the so-called RG functions (the $\beta$ and $\gamma$ functions) are given as follows:
\begin{eqnarray}
\beta_g &\equiv& \mu \partial_{\mu}g=g(-2\varepsilon + 3\gamma_1), \label{BetaG} 
\\
\beta_u &\equiv& \mu \partial_{\mu}u=u(\gamma_1-\gamma_2), \label{BetaU}
\\
\gamma_{i} &\equiv& \mu \partial_{\mu} \ln Z_i,\quad i=1,2,\label{GammaS}
\end{eqnarray}
and are based on relations among the renormalization constants (\ref{Renorm12}) together with explicit expressions of $Z_1$ and $Z_2$ given by (\ref{Z1expansion}) and (\ref{Z2expansion}), respectively. To obtain the IR asymptotic behavior of the correlation functions deep inside of the inertial interval we need to identify the coordinates ($g^∗$, $u^∗$ ) of the corresponding IR stable fixed point where $\beta_g$ and $\beta_u$ vanish, i. e.:
\begin{equation}
\beta_g(g_{*})=0, \quad \beta_u(g_{*},u_{*})=0, \label{BetaZero}
\end{equation}
where $g_{*}\neq 0$ and $u_{*}\neq 0$ in two loop approximation are required to have the form
\begin{eqnarray}
g_{*}&=&g_{*}^{(1)} \varepsilon + g_{*}^{(2)} \varepsilon^2 + O(\varepsilon^3), \label{Gexpansion} \\
u_{*}&=&u_{*}^{(1)}+u_{*}^{(2)} \varepsilon + O(\varepsilon^2). \label{Uexpansion}
\end{eqnarray}
It may be verified by direct calculation that at non-trivial fixed points the following expressions hold:
\begin{eqnarray}
g_{*}^{(1)}&=&\frac{(2\pi)^d}{S_d}\frac{8(d+2)}{3(d-1)}, \label{Gstar1} 
\\
g_{*}^{(2)}&=&\frac{(2\pi)^d}{S_d}\frac{8(d+2)}{3(d-1)} \lambda, \label{Gstar2} 
\\
u_{*}^{(1)}&=&\frac{1}{3a_2}\left( -2a_2 - \frac{\sqrt[3]{2} b_1}{\sqrt[3]{b_2 + b_3}} + \frac{ \sqrt[3]{b_2 + b_3}}{\sqrt[3]{2}}  \right), \label{Ustar1} 
\\
u_{*}^{(2)}&=&\frac{2(d+2)}{d [1+2 u^{(1)}_{*}]} \Biggl[\lambda - \frac{128 (d+2)^2}{3(d-1)^2} \mathcal{B}(u^{(1)}_{*}) \Biggr],  \label{Ustar2}
\end{eqnarray}
where $\lambda$ is related to the coefficient $z_{21}^{(1)}$ in Eq.\,(\ref{Z1expansion}) as
\begin{equation}
\lambda = \frac{2}{3} \frac{(2\pi)^{2d}}{S_d^2}\left(\frac{8(d+2)}{d-1}\right)^2 z_{21}^{(1)}. \label{Lambda}
\end{equation}
Coefficient $\mathcal{B}(u^{(1)}_{*})$ is discussed in the text below. Let us now give the explicit expressions for 
$a_i$ with $i \in 0, 1, 2$ and $b_i$ with $i \in 1, 2, 3$. They read:
\begin{eqnarray}
b_1 &=& a_2 \left( 3a_1 - 4a_2\right) \label{b1}
\\
b_2 &=& a_2^2 \left(-27 a_0 + 18 a_1 -16 a_2 \right) \label{b2}
\\
b_3 &=& \sqrt{4 b_1^3 + b^2_2} \label{b3}
\\
a_0 &=& 2 \left[ d^2 -3 + A(A+d)\right] \label{a0}
\\
a_1 &=& 6 (1-A^2) - 2 A (d-2) - d(d+1) \label{a1}
\\
a_2 &=& d(d-1) \label{a2}
\end{eqnarray}
The value of the coefficient $a_1$ differs from that presented in Ref.\,\cite{Jurcisin2016} where most probably a typesetting error occurred since all further results of Ref.\,\cite{Jurcisin2016} agree with corresponding results of the present helical $A$ model when the limit $\rho \rightarrow 0$ is taken. Moreover, $a_1$ presented in Ref.\,\cite{Jurcisin2016} takes the same form as the current one when the (probably misplaced) brackets are corrected. 

As already mentioned, one loop results given by Eqs.\,(\ref{Gstar1}) and (\ref{Ustar1}) are free of helical contributions. Further, $g^{(2)}_{*}$ depends exclusively on the properties of the underlying velocity field which in turn means that it is common within a class of models with passively advected admixtures as discussed for example in Ref.\,\cite{Jurcisin2014}. In more detail, $g^{(2)}_{*}$ is completely determined by $\lambda$ from Eq.\,(\ref{Lambda}) and takes exactly the same value as the corresponding quantity in Refs.\,\cite{Jurcisin2016}. However, $u^{(2)}_{*}$ is model specific and known only for special choices of $A \in  0, 1 $, see Ref.\,\cite{Jurcisin2014} for more details. Here, it is expected to contain helical contributions via the quantity $\mathcal{B}(u^{(1)}_{*})$ which in turn is completely given by the coefficient $z_{21}^{(2)}$ in Eq.\,(\ref{z2_21helical}) and it obtains the following value at $u = u^{(1)}_{*}$ :
\begin{eqnarray}
\mathcal{B}(u^{(1)}_{*},\rho ) = \frac{(2\pi)^{2d}}{S^2_d} \, z^{(2)}_{21} (u_{*}^{(1)},\rho) \label{Bstar}
\end{eqnarray}
We retained the $d$ dependencies for notation purpose. However as already mentioned, only spatial dimension $d = 3$ is physically meaningful when helical effects are considered. The IR behavior of the fixed point is determined by the matrix of the first derivatives which is given as
\begin{equation}
\Omega_{ij} = \left(
\begin{array}{cc}
\partial \beta_g / \partial g  & \partial \beta_g / \partial u  \\
\partial \beta_u / \partial g  & \partial \beta_u / \partial u   \\
\end{array}
\right) \label{IRmatrix}
\end{equation}
and is evaluated for given ($g_{*}$, $u_{*}$). The present matrix has a triangular form since $\beta_g$ is independent of $u$ and thus $\partial \beta_g/ \partial u =0$. Thus, diagonal elements $\partial \beta_g / \partial g$ and $\partial \beta_u / \partial u$ correspond directly to the eigenvalues of the present matrix. Subsequently, using numerical analysis one can show that real parts of diagonal elements are positive for all values of $A$ in vicinity of $\epsilon =0$. Furthermore, we have aslo shown that including spatial parity violation shifts the values of the present matrix even further to positive values. In the end, we stress out the well known fact that $\beta$ functions of the present model are exactly given even at the one-loop order since all higher order terms cancel mutually which means that the anomalous dimensions $\gamma_1 = \gamma_2$ equal exactly $2\varepsilon/3$ at the IR stable fixed point.

\section{Helicity and the turbulent Prandtl number} \label{sec5}

As discussed in the text above, all one loop contributions to the renormalization constants $Z_1$ and $Z_2$ are free of helical contributions even when turbulent environments with broken spatial parity are considered explicitly \cite{Jurcisin2016,Adzhemyan2005}. In the previous section, we have therefore determined two loop values of renormalization constants $Z_1$ and $Z_2$ which in fact do manifest helical effects for both renormalization constants. Additionally, stable non-trivial IR fixed point is shown to exists for given $g^*$ and $u^*$ in two loop order of calculation. Therefore, one may expect (as already seen for example in Refs.\cite{Jurcisin2014,Jurcisin2016}) that two loop order is sufficient to capture the leading order helical contributions to the required turbulent Prandtl number which then of coarse correspond to the two loop order of given perturbative theory. We prove this assertion in the subsequent text by explicit determination of the corresponding values of turbulent Prandtl number for a range of values of the continuous parameter $A$. However, we show explicitly that some regions of $A$ have to be omitted when spatial parity violation is weak enough.

Two-loop calculation presented here is to a large extent based on Ref.\,\cite{Adzhemyan2005} where the turbulent Prandtl number in the simple model of passive advection of a scalar field has been calculated for completely symmetrical turbulent environment. As shown for example in Ref.\,\cite{Jurcisin2014}, the approach of Ref.\,\cite{Adzhemyan2005} may successfully by used also in helical environments. Moreover, although tensorial properties of passively advected fields considered in Refs.\,\cite{Jurcisin2014,Adzhemyan2005,Jurcisin2016} are to a large extent different from those considered in the present model, one further analogy is observed when the corresponding correlators of stochastic pumping are reviewed. Therefore, the actual calculations of two loop Prandtl number performed here are closely analogous to those of Refs.\,\cite{Jurcisin2014,Adzhemyan2005} and the resulting two loop expression for the Prandtl number is analogous to Eq.\,(33) of Ref.\,\cite{Adzhemyan2005}. Moreover, due to the passive nature of the admixtures considered here and in Refs.\,\cite{Jurcisin2014,Jurcisin2016,Adzhemyan2005}, we note that all (partial) results which depend not on the admixture field $\mathbf b$ have to be identical in Refs.\,\cite{Jurcisin2014,Jurcisin2016,Adzhemyan2005} as well as in the present model. In explicit, properties of the helical environments with given admixture type which fully correspond to the model given by differential Eqs.\,(\ref{BB}) and (\ref{vv}) are in two loop order of the perturbation theory of the corresponding field theoretic model completely encoded by the Feynman graphs of Fig.\,\ref{fig3}.

Taking together, although present calculations are analogous to that of Refs.\,\cite{Jurcisin2014,Adzhemyan2005}, all quantities inherently connected with given admixtures and their interactions have to be reexamined here. We also stress out that formula (33) of Ref.\,\cite{Adzhemyan2005} holds inside of the inertial interval and does not depend on the renormalization scheme. The details of the calculations are outlined in Ref.\,\cite{Adzhemyan2005} and we omit them consequently. The resulting two loop expression for Prandtl number is obtained as
\begin{eqnarray}
u_{eff}&=& u^{(1)}_{*} \Biggl(1 + \varepsilon \Biggl\{ \frac{1+u^{(1)}_{*}}{1+2 u^{(1)}_{*}} \Biggl[\lambda - \frac{128 (d+2)^2}{3(d-1)^2} \mathcal{B}(u^{(1)}_{*}) \Biggr] \nonumber 
\\ 
&& + \frac{(2\pi)^d}{S_d}\frac{8 (d+2)}{3(d-1)}\left[a_{v}-a_{b}(u^{(1)}_{*})\right]\Biggl\} \Biggr), \label{InversePrandtl}
\end{eqnarray}
where $\varepsilon$ and dimension $d$ are taken to their physical values of $\varepsilon = 2$ and $d = 3$, the one loop value of turbulent Prandtl number $u_{*}^{(1)}$ has already been given in Eq.\,(\ref{Ustar1}), $\mathcal{B}(u^{(1)}_{*})$ is given in Eq.\,(\ref{Bstar}) and $\lambda$ in Eq.\,(\ref{Lambda}). The following numerical value corresponds to $\lambda$ in $d = 3$ as considered here for helical environments:
\begin{equation}
\lambda = -1.0994 - 0.0556 \times 10^{-3} \rho^2 \label{LambdaNumerical}
\end{equation}
which is the same as in Ref.\,\cite{Jurcisin2014} since $\lambda$ in independent of the admixture type for passive advection. The remaining parameters $a_v$ and $a_b$ which enter Eq.\,(\ref{InversePrandtl}) are discussed in the text below. Let us first notice that $a_v$ and $a_b$ represent the finite parts of one-loop diagrams with two external velocity type fields $\mathbf v$, $\mathbf v^{\prime}$ and two admixture type fields $\mathbf b$, $\mathbf b^{\prime}$ respectively. Since turbulent velocity environments here and in Ref.\,\cite{Adzhemyan2005} are the same, the coefficient $a_v$ must be also the same and we shall not reproduce its analytic form here. In $d = 3$, it can however be easily evaluated numerically as
\begin{equation}
a_v = -0.047718/(2\pi^2). \label{AvNumerical}
\end{equation}
Contrary, $a_b$ is model specific and is given by the finite part of one-loop one irreducible diagram $\Gamma^{(1)}$  making it thus also $\rho$ independent. As already discussed, the present generalized helical $A$ model and the less general model introduced in Ref.\,\cite{Jurcisin2016} have all one-loop quantities including $a_b$ identical due to helical effects beeing pronounced first in the two-loop order. Since $a_b$ plays a crucial role in two-loop calculation of inverse turbulent Prandtl number that follows show it explicitly in the present paper. After straightforward calculation discussed for example in Ref.\,\cite{Adzhemyan2005}, one obtains $a_b$ in the same form as authors of Ref.\,\cite{Jurcisin2016}. It reads:
\begin{widetext}
\begin{eqnarray}
a_b(u) &=&-\frac{S_{d-1}}{2 u (d-1) (2\pi)^d}\int_0^{\infty} dk \int_{-1}^1 dx\, (1-x^2)^{\frac{d-1}{2}} \nonumber 
\\
&\times& \left\{
\frac{k \left[ k^3 x A (1-A) + k^2 ( x^2 (1-A^2)+A+d-2) +2 k x (d-1)+d-1 \right]}{  (k^2+2 k x +1) \left[ (1+u)k^2+2 u k x + u \right]} \right. \nonumber
\\
&-& \left. \frac{ \theta(k-1) \left[ k A (1-A)(1+u)x +A^2(1+3u)x^2 +A (1+u-2(1+2u)x^2) +(1+u)(x^2+d-2) \right] }{k(1+u)^2} \right\} \label{Ab}
\end{eqnarray}
\end{widetext}
\begin{figure*}
\vspace{0.5cm}
\begin{center}\includegraphics[width=15cm]{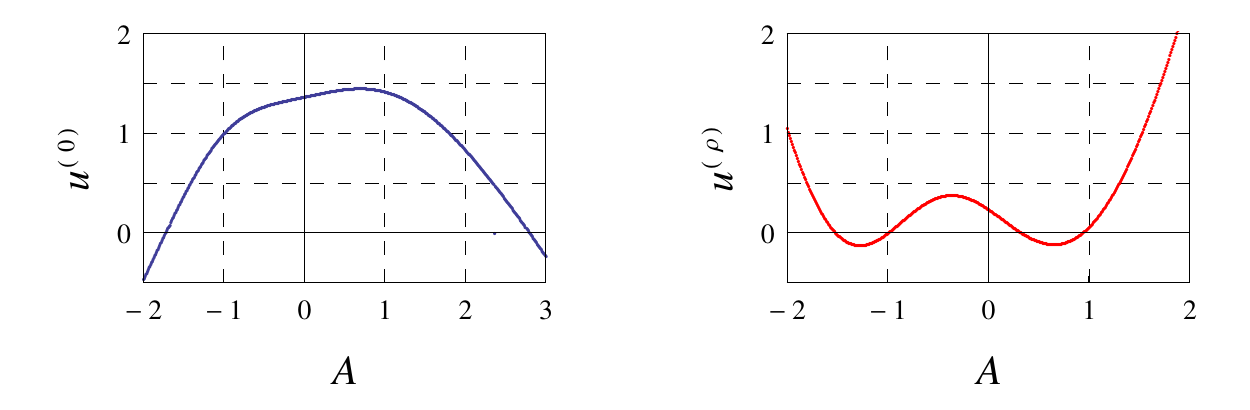}\end{center}
\caption{ (Color online) Dependence of $u^{(0)}$ and $u^{(\rho)}$ on parameter $A$ shown in regions $-2 \leq A \leq 3$ and $-2 \leq A \leq 2$ respectively. Quantity $u^{(0)}$ corresponds to non-helical value of inverse Prandtl number, while $u^{(\rho)}$ represents helical contribution to the inverse Prandtl number. Points represent numerical values obtained from Eq.\,(\ref{Split}).
\label{fig5}}
\end{figure*}

\noindent with $\theta(k-1)$ being the usual Heaviside step function with $k-1$ as argument. The expression (\ref{Ab}) is be easily obtained by direct calculation of the single one loop diagram shown in Fig.\,\ref{fig3}. One further difference manifested even at one loop order lies in the already calculated value of $u_{*}^{(1)}$. Due to the tensorial  interaction structures in the present model it obtains the form of Eq.\,(\ref{Ustar1}) which is of coarse different from the corresponding value obtained in Ref.\,\cite{Adzhemyan2005}. Using the expression (\ref{InversePrandtl}) with all necessary coefficients now known due to Eqs.\,(\ref{Ustar1}), (\ref{Lambda}), (\ref{Bstar}), (\ref{AvNumerical}) and (\ref{Ab}), we may proceed to the actual calculation of the Prandtl number. As already discussed, using RG techniques in theory of critical behavior requires to substitute $\varepsilon = 2$ in the final expressions as thoroughly discussed for example in Refs.\,\cite{VasilevBook,McComb}. The spatial dimension is set to $d=3$ as required by the nature of the helical problem. Inserting all necessary quantities into the Eq.\,(\ref{InversePrandtl}) we obtain its values for arbitrary $A$. In other words, we get the inverse turbulent Prandtl number $u_{eff}$ as a function of $A$ which we indicate here explicitly by denoting the corresponding values as $u_{eff} (A)$. 

Since the corresponding Eqs.\,(\ref{InversePrandtl}), (\ref{Ustar1}), (\ref{Lambda}), (\ref{Bstar}), (\ref{AvNumerical}) and (\ref{Ab}) are all known in analytic form also the resulting turbulent inverse Prandtl number has an analytic form. Nevertheless, due to the complicated analytic structure of the coefficient $z^{(2)}_{12}$, the expression in Eq.\,({\ref{Bstar}}) is a complicated analytic function of model variables. Thus, the resulting analytic expression for the inverse Prandtl number is also lengthy and complicated. Consequently, we shall not show it here explicitly (all necessary coefficients for it's calculation are discussed either in the main body of the article or in its Appendix) but instead we split $u_{eff} (A)$ into its non-helical part $u^{(0)} (A)$ and its corresponding  helical contribution $\rho^2 u^{(\rho)} (A)$ in the following way: 
\begin{equation}
u_{eff} (A) = u^{(0)} (A) + \rho^2 u^{(\rho)} (A). \label{Split}
\end{equation}
Note that both $u^{(0)} (A)$ and $u^{(\rho)} (A)$ are defined to be independent of $\rho$ but $u^{(\rho)} (A)$ is the coefficient which stands in front of the helical contribution in Eq.\,(\ref{Split}) and encodes thus all helical effect of the present model. As before, both $u^{(0)} (A)$ and $u^{(\rho)} (A)$ are quite complicated analytic functions of model parameters and we therefore present them here via their graphical representation given in Fig.\,\ref{fig5} which is sufficient for the interpretation of the obtained results. Moreover, the corresponding numerical values are given in Tab.\,\ref{tab1} for few selected values of parameter $A$. In Fig.\,\ref{fig5}, we plot $u^{(0)} (A)$ in the region of $-2 \leq A \leq 3$ as it contains all zero points of the present function. Due to the same reasoning, $u^{(\rho)} (A)$ is plotted in a smaller region of $-2 \leq A \leq 2$. The actual turbulent Prandtl number $Pr_A$ is then given as the inverse of $u_{eff} (A)$. In explicit:
\begin{equation}
Pr_A = \frac{1}{u^{(0)} (A) + \rho^2 \, u^{(\rho)} (A)}. \label{Prandtl}
\end{equation}
However, in the immediately following text we shall rather use the corresponding values of the inverse turbulent Prandtl number as they better suit our next discussion. Afterwards, we discuss turbulent Prandtl numbers in helical $A$ model for selected values of $A$. 

Let us now therefore consider non-helical part $u^{(0)} (A)$ of the inverse turbulent Prandtl. First, we note that in the range $-1 \leq A \leq 1$ non-helical values of the function $u^{(0)} (A)$ obtained here are in complete agreement with those obtained by authors of Ref.\,\cite{Jurcisin2016} for a simpler non-helical case (see Fig.\,(8) in Ref.\,\cite{Jurcisin2016} and the corresponding analytic expressions in the Appendix of the same reference). However, in Ref.\,\cite{Jurcisin2016} only the region $-1 \leq A \leq 1$ is investigated and thus the important zero points of function $u^{(0)} (A)$ have not been discussed in any way. However, problems which arise at zero points of $u^{(0)} (A)$ clearly manifest the limits of perturbative two-loop approach as used here. Physically, when $u^{(0)} (A)=0$ the effective value of the corresponding diffusion coefficient for given $A$ should be infinitesimally small which is of coarse non-physical. Nevertheless, zero points of the function $u^{(0)} (A)$ are present and located at $A = -1.723$ and $A = 2.800$ (numerical values rounded on the last digit). Consequently, by approaching the zero points of $u^{(0)}$, turbulent Prandtl numbers would obtain infinitely large values. Additionally, according to Fig.\,\ref{fig5}, inverse turbulent Prantdl number would be negative in regions $A < -1.723$ and $A > 2.800$. The effective diffusion coefficients would in such cases obtain non-physical values which clearly must be avoided. 

Thus, for non-helical turbulent environments constraints $-1.723 \leq A \leq 2.800$ must be imposed on values of $A$ in the two loop order of perturbation theory. Additionally, values of $A$ close to zero points of $u^{(0)} (A)$ should also be considered only with extreme caution as the resulting turbulent Prandtl numbers tend to $+\infty$ at the border of the allowed interval. On the other hand, such a problem did apparently not occur for the corresponding one loop values as clearly demonstrated in Fig.\,\ref{fig4} in the present paper. It is therefore clear, that constraints for non-helical environments arise only in connection with the two loop order calculation used here and are therefore inherently given by the structure of perturbation theory of the $A$ model. In other words, such constraints are not inherent to values of $A$ outside of the usually studied region $-1 \leq A \leq 1$ and represent only an artifact of the perturbative approach. Such a conclusion is supported also by the special case of $A = -1$ discussed later in more detail. For now on, we stress out that all previous conclusions are completely true only in non-helical environments. Bearing in mind the constraints on $A$ in non-helical case, we also notice that $u^{(0)} (A)$ has a maximum at $A=0.7128$ (rounded on last presented number) and is quite well stable in the range of approximately $-0.5<A<1.5$ which in connection to the results on one-loop order values presented for example in Fig.\,\ref{fig4} also explains the remarkable stability of models with $A=0$ and $A= 1$ against the order of perturbation theory as already noticed in Ref.\,\cite{Jurcisin2016}. Qualitatively similar picture holds also when helical contributions are considered as discussed below.

Let us now finally turn our attention to $u^{(\rho)}$ which encodes the much needed helical contributions of our generalized helical $A$ model. Its graphical representation is given in Fig.\,\ref{fig5} and is directly connected with the inverse Prandtl number $u_{eff} (A)$ by Eq.\,(\ref{InversePrandtl}). Consequently, the sign of $u^{(\rho)}$ determines the character of helical dependence of $u_{eff} (A)$. In explicit, for positive (negative) values of $u^{(\rho)} (A)$ the corresponding inverse turbulent Prandtl number will be a monotonically growing (descending) function of helicity parameter $\rho$. The zero points of $u^{(\rho)} (A)$ turn out therefore to represent very important special cases of the general $A$ model. Their location is easily determined numerically based on the previous analysis with resulting values being $-1.516$, $-1.000$, $0.325$ and $0.912$ (numbers rounded at the last presented digit). 

\begin{figure}[t]
\begin{center}\includegraphics[width=7cm]{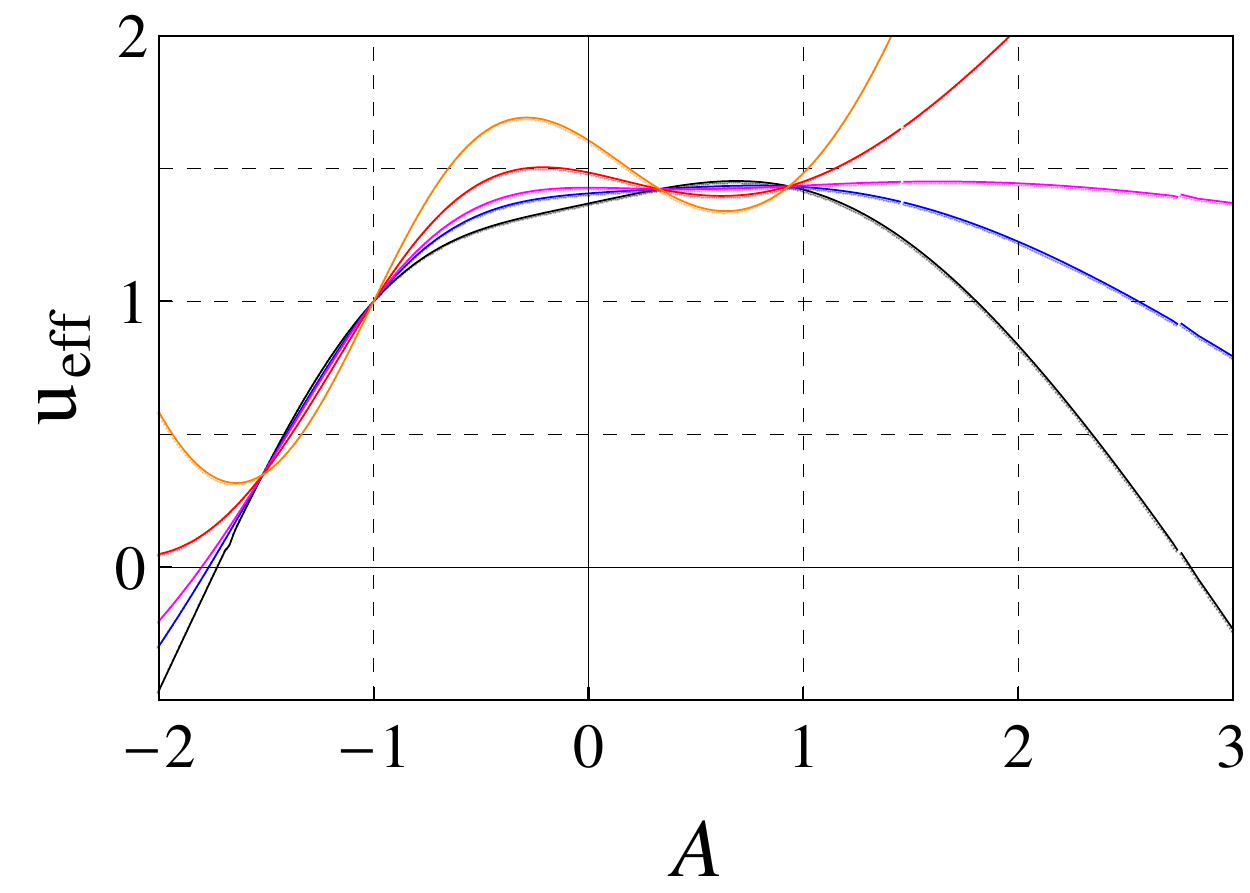}\end{center}
\caption{ (Color online) Inverse turbulent Prandtl numbers $u_{eff}$ as function of $A$ shown in the range of $-2 \leq A \leq 3$ for selected values of helical parameter $\rho$. Shown are values of $u_{eff}$ for $\rho=0$ (black), $\rho=0.4$ (blue), $\rho=0.5$ (magenta), $\rho=0.7$ (red) and $\rho=1$ (orange). \label{fig6} }
\end{figure}

Furthermore, inserting values of functions $u^{(0)}(A)$ and $u^{(\rho)}(A)$ into Eq.\,(\ref{InversePrandtl}) one may easily calculate the inverse turbulent Prandtl number $u_{eff}(A)$ as a function of $A$ for selected values of $\rho$. The resulting values are presented in Fig.\,\ref{fig6} and show highly interesting behavior. In non-helical case, the resulting turbulent Prandtl numbers have been shown to obtain unphysical values in restricted intervals $A < -1.723$ and $A > 2.800$. However, $u^{(\rho)}(A)$ is according to Fig.\,\ref{fig5} in both restricted intervals not only positive but it also evidently satisfies $u^{(\rho)}(A)>|u^{(0)}(A)|$. Therefore, when exceeding some critical value of helicity parameter $\rho$ for given value of $A$ the corresponding inverse turbulent Prandtl number $u_{eff}=u^{(0)} +\rho^2 u^{(\rho)}$ must get positive. In other words, when parity violation is strong enough, the resulting inverse turbulent Prandtl number obtains always positive values. Thus, introducing parity violation into the turbulent system improves perturbative series for the present model as shown explicitly in Fig.\,\ref{fig6}. In this respect, we also notice that increasing $\rho$ from $0$ up to $\rho\approx0.5$ enlarges the region of physically allowed values of $A$. However, the allowed region of $A$ grows according to Fig.\,\ref{fig6} infinitely when helicity parameter $\rho$ is increased further. Strikingly, it is not required to reach the maximum possible violation of parity ($|\rho|=1$) to remove the constraints on $A$. Contrary, by exceeding a critical value of $\rho= 0.749$ (rounded on the last presented digit) we remove any constraints on $A$ completely. In other words, beyond the critical value of $\rho =0.749$ all inverse turbulent Prandtl numbers are positive and thus physical. Consequently, exceeding the threshold of $\rho=0.749$ stabilizes the diffusion advection processes in the general $A$ model to a large extent. Calculations within the two loop order of the corresponding perturbative theory are then well defined which even further our hypothesis regarding the artificiality of constraints imposed on values of $A$ in non-helical environments. The interplay between the interaction parameter $A$ and parameter $\rho$ describing the amount of spatial parity violation is thus proven to be highly non-trivial. 

Additionally, in Fig.\,\ref{fig6} we may identify values of $A$ for which the helical dependence of inverse turbulent Prandtl number is relatively small. Such regions are all connected with the regions of negative values of $u^{(\rho)}$ and correspond therefore to the union of interval $-1.516 \leq A \leq -1.000$ with $0.325 \leq A \leq 0.912$. Interestingly, we notice that two special cases $A=-1$ and $A=1$ lie either directly in such regions ($A=-1$ case) or are located in a close vicinity of these ($A=1$ case). First, let us discuss the case of linearized helical NS equations with $A = -1$ which up to date has not been investigated in any way. According to the performed numerical analysis of Eq.\,(50), $u^{(\rho)}$ is less than $10^{-8}$ at $A = -1$ which in limits of accuracy means that $u^{(\rho)}$ is actually equivalent to zero and consequently $A = -1$ corresponds directly to the zero point of $u^{(\rho)} (A)$. We stress out that this is not just a trivial influence of vanishing of all helical terms in two-loop diagrams $\Gamma_l^{(2)}$ with $l=1, \ldots, 8$. In fact, separately each diagram contains corresponding helical terms which however mutually cancel each other when all diagrams are summed up together as required in deriving of $\Sigma^{b^{\prime}b}$. As a consequence, at $A = -1$ the properties of the flow are completely independent of spatial parity violation of the underlying fully developed turbulent velocity flow. Moreover, this result is most probably independent of perturbation order as suggested by $u^{(0)}$ being exactly one (within the accuracy of the present numerical analysis) at both the first and the second order of the corresponding perturbation order. A similar hypothesis has already been stated by authors of Refs.\,\cite{Jurcisin2016} for non-helical values. Here, we however demonstrate that such a behavior persist even in helical environments.

In this respect, it is also worth to mention that for $A = 1.038$ (value obtained numerically and rounded on the last presented digit) one and two loop values of non-helical inverse Prandtl number do also coincide, a result which was not observed in Ref.\,\cite{Jurcisin2016} due to constraining the analysis only at $-1 \leq A \leq 1$. However, unlike for the $A = -1$ case, helical effects are present quite significantly for the $A = 1.038$ case (as later discussed more closely, the difference between the non-zero value of the turbulent Prandtl number and its minimal value at $|\rho|=1$ is around $7 \%$.) which means that the model of linearized Navier-Stokes equations corresponding to the $A = -1$ case in the present model has unique features. The remaining three zero points of $u^{(\rho)}(A)$, namely $A \in \{ -1.516, 0.325, 0.912 \}$, do not show the same behavior. Instead, their one and two loop values differ significantly. In other words, although the remaining three zero points of $u^{(\rho)} (A)$ also lead to models stable against helical effects in two loop order, there is no indication that higher order of perturbation theory preserve location of the zero points for the analog of $u^{(\rho)}(A)$ calculated in higher orders. 

\begin{table*}[t]
\begin{tabular}{ |c |c |c |c |c |c |c |c |c |c |c |c | }
\hline
$A$             &   $-2$    & $-1.5$   &$-1.0$   & $-0.5$   &   $0$    &  $+0.5$   &  $+1.0$   &    $+1.5$ & $+2.0$  & $+2.5$ & $+3.0$
\\ \hline
$u^{(0)} (A)$   & $-0.4663$ & $+0.3726$ & $+1.000$ & $+1.2705$ & $+1.3685$ & $+1.4436$ & $+1.4205$ & $+1.2145$ & $+0.8343$ & $+0.3339$ & $-0.2355$
\\ \hline
$u^{(\rho)}(A)$ & $+1.0503$  & $-0.0163$ & $-0.000$ & $+0.3587$ & $+0.2376$ & $-0.0854$ & $+0.0623$ & $+0.9444$ & $+2.4408$ & $ +4.3269$ & $+6.4228$
\\ \hline
\end{tabular}
\caption{Turbulent Prandtl number for the present helical model is given as $Pr_A = 1 / (u^{(0)} (A) + \rho^2 \, u^{(\rho)} (A) )$ with numerical values of $u^{(0)} (A)$ and $u^{(\rho)} (A)$ given for selected values of $A$ and rounded on the last presented digit. Values at $A=-2$, $A=2.5$ and $A=3$ demonstrate physical constraints that must be imposed on values of $A$ in two loop calculations as performed here. In fact, for non-helical case only values satisfying $-1.723 < A < 2.800$ are considered in the present paper. Let us however note that for sufficiently large values of $\rho$, namely $\rho < 0.749$ all turbulent Prandtl number obtain positive and thus physically meaningful values. \label{tab1}}
\end{table*}

On the other hand, the equality of one and two loop results for $A = 1.038$ explains another up to date not well understood result of Ref.\,\cite{Jurcisin2014}. Here, the authors have observed that kinematic MHD model corresponding to $A=1$ of the present model is remarkably stable against one- and two-loop order corrections. Using however the previous result we easily explain this as a consequence of $A=1$ case lying in the proximity of $A = 1.038$ where one- and two-loop order values are identical. Such a situation is of coarse true only in the present two-loop order of the calculation. In higher orders of the perturbation theory, the corresponding polynomials over $A$ which occur in diagrams of Fig.\,\ref{fig3} are of higher orders and consequently the intersection between higher order analog of $u^{(0)} (A)$ and one-loop order result $u^{(1)}_{*}$ may dramatically shift to new values. Additionally, contrary to the $A=-1$ case, there is no evidence from helical values that the location of $A = 1.038$ would be fixed in higher order loop calculations.  Thus, the relatively small contribution of the two-loop order corrections to the inverse Prandtl number of the kinematic MHD model should be clearly attributed to the present two-loop order of calculation. Additionally to this, $A=1$ case lies close to the border of the interval $A>0.912$ with $A=0.912$ being value where inverse turbulent Prandtl number is independent of any helical effects. Since the function $u^{(\rho)} (A)$ is continuous it consequently causes the helical effects for all models in vicinity of the $A=0,912$ case to be relatively well stable against helical effects. Such an effect is now clearly manifested also in the kinematic MHD model corresponding to $A=1$ case of the present model. Here, the corresponding inverse turbulent magnetic Prandtl number changes less than $5\%$ of its original non-helical value as observed for example in Ref.\cite{Jurcisin2014}.

Contrary to the previous two special cases discussed above, another physically important model corresponding to the $A = 0$ case of the present model lies deep in the interval of positive values of $u^{(\rho)} (A)$ and is thus located far away form points $A=0.325$ and $A=0.912$ where the function $u^{(\rho)} (A)$ has its zero points located. On the other hand, the $A=0$ model as studied for example in Ref.\,\cite{Jurcisin2014} lies relatively closely to the local maximum of the function $u^{(\rho)} (A)$ on the interval of $0.325 \leq A \leq 0.912$. Consequently, helical effects in the $A=0$ model are pronounced far greatly (almost by maximum possible amount in the interval of positive values of $u^{(\rho)}(A)$) as seen for example on the almost $20 \%$ change of the inverse turbulent Prandtl number in he helical environments. In other words, function $u^{(\rho)} (A)$ represents an easy tool to asses the importance of helical effects for given values of $A$ in the present model and explains previously unidentified context between the $A=0$ and $A=1$ models.

\begin{figure}[t]
\begin{center}\includegraphics[width=7cm]{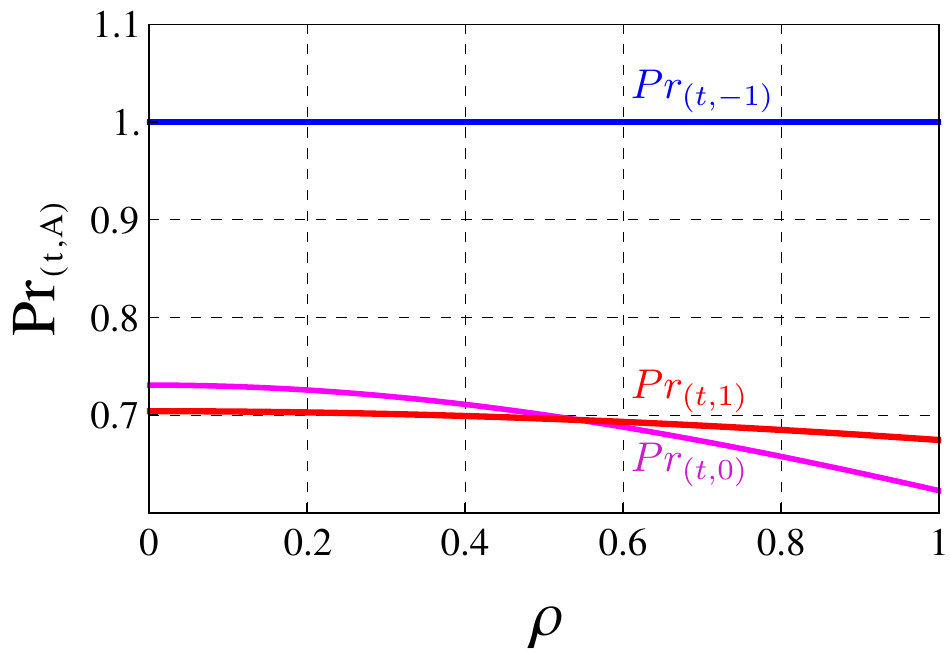}\end{center}
\caption{ (Color online) Helical dependence of turbulent Prandtl numbers $Pr_{t,A}$ for three physically important models with $A \in \{ -1,0,1\}$ shown in the range of $\rho \leq 1$. Presented curves correspond to values $A=-1$ (blue), $A=0$ (magenta) and $A=1$ (red). \label{fig7} }
\end{figure}

Finally, let us discuss the obtained values of helical turbulent Prandtl numbers which follow from Eq.\,(\ref{InversePrandtl}) and the functions $u^{(0)}$ and $u^{(\rho)}$ which appear therein. For selected parameters $A$, we show their corresponding numerical values in Tab.\,\ref{tab1} while their graphical representation is given in Fig.\ref{fig7} for the three physically important models of $A \in \{ -1, 0, 1\}$. As before, function $u^{(\rho)}(A)$ encodes the behavior of turbulent Prandtl numbers in respect to $\rho$ for all physically admissible values of $A$ which as shown before clearly depend also on the helicity parameter $\rho$. For the case of turbulent Prandtl numbers it has according to the Eq.\,(\ref{Prandtl}) the following meaning: For $u^{(\rho)}(A)>0$ turbulent Prandtl number is a decreasing function of $\rho$, for $u^{(\rho)}(A)<0$ turbulent Prandtl number is an increasing function of $\rho$ and finally for $u^{(\rho)}(A)=0$ turbulent Prandtl number is independent of $\rho$. This means that turbulent Prandtl number do increase with helicity parameter $\rho$ only for values of $A$ satisfying $-1.516 < A < -1$ and $0.325 < A< 0.912$. Excluded the zero points of $u^{(\rho)}(A)$, the remaining values of $A$ lead to monotonically decreasing helical turbulent Prandtl numbers as already seen in Ref.\,\cite{Jurcisin2014} for the special cases $A=0$ and $A=1$. While for $A=0$ model helical effects are pronounced more effectively due to reasons discussed above, the turbulent Prandtl number for $A=1$ model corresponding to kinematic MHD model is less sensitive to helical effects due to its above discussed proximity to the $A=0.912$ case. We also stress out, that there are no restrictions on $A$ when the threshold of $\rho \approx 0.749$ is exceeded. Thus, corresponding helical dependences of turbulent Prandtl numbers may for $\rho > 0.749$ also be considered. Consequently, we see that not the internal vectorial nature of the admixture itself but their interactions with the underlying turbulent field $\mathbf v$, as described by the parameter $A$, are crucial for developing different patterns in regard to helical effects and their influence on diffusion advection processes.

Taking together, we have shown that the impact of the interactions as given via the parameter value of $A$ has a highly non-trivial impact on diffusion-advection processes when helical turbulent environments are considered. The resulting dependencies are truly complicated functions of $A$ and lead to non-trivial effects in connection with helicity parameter $\rho$. Therefore, instead of the tensorial nature of the admixture itself we have clearly identified the tensorial structure of interactions to be a more dominant factor which effectively alters advection diffusion process in fully developed turbulent environments. Thus, assertions made by authors of Ref.\,\cite{Jurcisin2014} must partially be revided at least for the case of vector admixtures advected passively in turbulent environments and the greater than expected impact of interactions on the actual advection diffusion processes must be recognized. Additionally, we once again stress out that present calculations clearly demonstrate that helical effects excert stabilizing effect on diffusion advection processes.

\section{Conclusion} \label{sec6}

Using the field theoretic renormalization group technique in the two-loop approximation, we have obtained analytic expressions for turbulent Prandtl number within the general $A$ model of passively advected vector impurity. Compared to Ref.\,\cite{Jurcisin2016} a more realistic scenario with effects of broken spatial parity has been considered by defining appropriate correlators of stochastic driving forces. Technically, the presence of broken spatial parity is described by helicity parameter $\rho$ ranging from $|\rho| = 0$ (no parity breaking) to $|\rho|=1$ (highest possible violation of spatial parity). Since our general helical $A$ model encompasses the less general non helical model of Ref.\,\cite{Jurcisin2016} we have been able to recover the results of Ref.\,\cite{Jurcisin2016} within the present calculations. However, the parameter $a_1$, in Eq, (46) has been shown to differ from the corresponding non helical value of Ref.\,\cite{Jurcisin2016}. However since further results show no differences and only the parameter a 1 from Ref.\,\cite{Jurcisin2016} is clearly not reproducing the well established results of Refs.\,\cite{Jurcisin2016}, we attribute the difference merely to a typographic error made by authors of Ref.\,\cite{Jurcisin2016}. 

Furthermore, additionally to helical effects we extended our study of the $A$ model to arbitrary real values of $A$ as suggested by Ref.\,\cite{Arponen2009} whereas in Ref.\,\cite{Jurcisin2016}, only the interval $-1 \leq A \leq$ 1 is considered. Nevertheless, although one loop values of physical quantities have all been shown to obtain meaningful values when passing to the two loop order we noticed negative values of turbulent Prandtl numbers for $A < -1.723$ and $A > 2.800$ (numbers rounded at the last presented digit) in non-helical case. Furthermore, we show that helical effects effectively enhance stability in the present model and lift of the restrictions imposed on $A$ when a critical threshold of $\rho \approx 0.749$ (rounded on the last presented digit) is exceeded. This points towards the conclusion that restrictions of $A$ to interval $-1.723 \leq A \leq 2.800$ are most probably only an artifact of two loop order perturbative calculations. Furthermore, in Sec.\,\ref{sec5} we have shown that Feynman diagrams corresponding to the $n$-th order of perturbation theory will generally have a form of polynomials in $A$ with the highest possible power of $A$ being $2n$. We therefore expect that higher orders of loop calculations shift or let even completely vanish all the zero points of inverse turbulent Prandtl number. Such a behavior has already been observed in one loop order for the analogous quantity $u^{(1)}_{*}$. Thus, it would be of high interest to go beyond the limits of two loop order, however such an analysis is technically demanding and beyond the scope of the present paper. Nevertheless, two loop order values obtained deep in the interval of $-1.723 \leq A \leq 2.800$ are clearly free of any problems which means that physically most interesting cases of $A \in \{-1, 0, 1\}$ can be safely considered at least in the two loop order of the perturbation theory. Thus, all restrictions on the values of $A$ should be considered as an artifact of the perturbative approach used in the present work.

For the case of the model of linearized Navier-Stokes equations ($A = -1$) we have obtained helical values of turbulent Prandtl number equal $1$ regardless of the presence of helical effects. It is therefore natural to expect that also higher orders of perturbation theory may preserve the same, a hypothesis already stated by authors of Ref.\,\cite{Jurcisin2016}. This adds another argument in favor of hypothesis that problems with range of physically admissible values of $A$ could be resolved completely in higher orders. Physically, the resulting values demonstrate remarkable stability of $A = -1$ case against helical effects. 

Effectively, the $A=1$ case corresponding to kinematic MHD model has been shown to have some similarities to $A=-1$ model with regard to its helical properties. Varying $A$ continuously allowed us to show that high stability of $A=1$ model is not due the vectorial nature of the admixture but due to its interactions given by $A=1$. Since it lies in the proximity of the $A=0.912$ case, where helical effects are not present in two-loop order, it must consequently be effectively less sensitive to helical effects. Contrary $A=0$ model is shown to lie far from values of $A$ where helical effects are not present. Consequently it shows significant dependence of turbulent Prandtl number on $\rho$.

Taking together, the case of $A = 1$ corresponding to the kinematic MHD turbulence, the case $A = 0$ model of passive vector admixture and the model of linearized Navier-Stokes equations have been brought into the context of the more general $A$ model. The interactions encoded by values of $A$ result in various patterns of behavior of turbulent Prandtl numbers. Thus, in regions of $1.516 \leq A \leq 1.000$ and $0.325 \leq A \leq 0.912$ (all numbers rounded on the last digit) the corresponding two loop turbulent Prandtl number are monotonically growing with $\rho$. Moreover, for values of $A \in \{ -1.516; -1; 0.325; 0.912\}$ turbulent Prandtl numbers are independent of $\rho$ but as previously discussed only $A=-1$ case is believed to retain this property also in higher order loop calculations. Finally, the remaining values of $A$ which belong to physically admissible region posses monotonically decreasing turbulent Prandtl numbers when $\rho$ is increased. We thus conclude that varying the interactions by changing the values of $A$ has a more profound effect on advection diffusion processes than the tensorial character of the admixture itself which significantly refines the conclusions made by authors of Ref.\,\cite{Jurcisin2016}.

\begin{acknowledgments}
The work was supported by VEGA Grant No.\,$1/0222/13$ of the Ministry of Education, Science, Research and Sport of the Slovak Republic. The authors gratefully acknowledge the hospitality of the Bogoliubov Laboratory of Theoretical Physics of the Joint Institute for Nuclear Research, Dubna, Russian Federation. P.Z. likes to express his gratitude to M.\,Dan\v{c}o and M.\,Jur\v{c}i\v{s}in for fruitful discussions which helped to carry out investigations presented here.
\end{acknowledgments}

\appendix

\section*{Appendix}\label{ApA}

Here, we present results on coefficient $B^{(\rho)}$. Let us define:
\begin{equation}
B^{(\rho)} = \sum\limits_{l=1}^{8} \, B_l^{(\rho)} \label{Structure}
\end{equation}
with $l = 1, \ldots, 8$ denoting separate contributions from two-loop diagrams labeled according to Fig.\,\ref{fig3}. Each of the coefficients $ B_l^{(\rho)}$ with $l \in 1, \ldots 8$ is a polynomial in $A$ of the order $4$ the most and depends on variables $u,d$ while $d$ in helical case is strictly equal to $3$. Graphs $\Gamma^{(2)}_l$ with $l \in 2, \ldots 8$ have been calculated using the technique presented in Ref.\,\cite{Adzhemyan2005} while the remaining $\Gamma^{(2)}_1$  has been calculated using the approach described in Ref.\cite{VasilevBook}. In this appendix, we show explicitly analytic expressions for graphs $\Gamma^{(2)}_1$ and $\Gamma^{(2)}_3$ but present the remaining six only graphically as the corresponding analytic expressions are extensive in their length. Let us start now with $B_1^{(\rho)}$ as its form is the most complicated. It reads:
\begin{equation}
B_1^{(\rho)} = 16u \left(A^2 f_1^{(2)} + A^3 f_1^{(3)} +A^4 f_1^{(4)}\right) \label{StructureGraph1} 
\end{equation}
where $A^i$ with $i \in 2, 3, 4$ are the corresponding powers of parameter $A$ while $f^{(i)}$ with $i \in 2,3,4$ are yet unspecified functions labeled by superscripts $(i)$. They read:
\begin{equation}
f_1^{(i)} = G_0^{(i)} (d,u) + \sum\limits_{j=1}^{11} \, g_j^{(i)} \, G_j(d,u) \label{Graph1general}
\end{equation}
Here, $j \in 1, \ldots 11$ runs over $11$ elements of the sum on the right hand side of Eq.\,(\ref{Graph1general}). Functions $G_j (d, u)$ carry no index $i$ and are consequently the same for all $i \in 2, 3, 4$. Contrary, functions $G_0^{(i)} (d,u)$ and $g_j^{(i)}$ depend on $i$ and shall be discussed later. Functions $G_j (d, u)$ read
\begin{eqnarray}
G_{1} (d,u) \, \, &=&\, _2F_1\left(-\frac{1}{2},+\frac{1}{2};\hspace{3mm} \frac{d}{2} ; \hspace{3mm}\frac{u^2}{(u+1)^2}\right), \label{FirstG}
\\
G_{2} (d,u) \, \, &=&\, _2F_1\left(+\frac{1}{2},+\frac{1}{2}; \hspace{3mm} \frac{d}{2} ; \hspace{3mm}\frac{u^2}{(u+1)^2}\right),  
\\ 
G_{3} (d,u) \, \, &=&\, _2F_1\left(-\frac{1}{2},-\frac{1}{2};\hspace{3mm} \frac{d}{2} ; \hspace{3mm}\frac{u^2}{(u+1)^2}\right), 
\\ 
G_{4} (d,u) \, \, &=&\, _2F_1\left(+\frac{1}{2},+\frac{3}{2};\frac{d+2}{2};\frac{u^2}{(u+1)^2}\right) 
\\ 
G_{5} (d,u) \, \, &=&\, _2F_1\left(-\frac{3}{2},-\frac{1}{2};\hspace{3mm} \frac{d}{2} ; \hspace{3mm}\frac{u^2}{(u+1)^2}\right), 
\\ 
G_{6} (d,u) \, \, &=&\, _2F_1\left(-\frac{1}{2},+\frac{1}{2};\frac{d+2}{2};\frac{u^2}{(u+1)^2}\right), 
\\ 
G_{7} (d,u) \, \, &=&\, _2F_1\left(+\frac{1}{2},+\frac{1}{2};\frac{d+2}{2};\frac{u^2}{(u+1)^2}\right), 
\\ 
G_{8} (d,u) \, \, &=&\, _2F_1\left(+\frac{1}{2},+\frac{5}{2};\frac{d+4}{2};\frac{u^2}{(u+1)^2}\right),
\\ 
G_{9} (d,u) \, \, &=&\, _2F_1\left(-\frac{1}{2},+\frac{3}{2};\frac{d+2}{2};\frac{u^2}{(u+1)^2}\right), 
\\ 
G_{10} (d,u) \, \, &=&\, _2\tilde{F}_1\left(+\frac{1}{2},+\frac{1}{2};\hspace{3mm} \frac{d}{2} ; \hspace{3mm}\frac{u^2}{(u+1)^2}\right), 
\\ 
G_{11} (d,u) \, \, &=&\, _2\tilde{F}_1\left(-\frac{1}{2},+\frac{3}{2};\frac{d+2}{2};\frac{u^2}{(u+1)^2}\right), \label{lastG}
\end{eqnarray}
where $_2 F_1$ stands an ordinary hypergeometric function and $_2 \tilde{F}_1$ for a regularized hypergeometric function. Furthermore, the $G_0^{(i)} (d, u)$ functions from (\ref{Graph1general}) are given as
\begin{eqnarray}
G_0^{(2)} &=& \frac{\pi^{3/2} (d-2) \left[ 3d(u+1)+u+9\right] \, \Gamma (\frac{d}{2})}{ 128 (d^2 +d-2)(u+1)^3 \Gamma (\frac{d+3}{2})},
\\
G_0^{(3)} &=& -\frac{\pi^{3/2} (d-2) \left[ (d-3)u+d+5\right] \, \Gamma (\frac{d}{2})}{ 64 (d-1)(d+2)(u+2)^3 \Gamma (\frac{d+3}{2})},
\\
G_0^{(4)} &=& -\frac{\pi^{3/2} (d-2) \left[ (d+7)u+d-1\right] \, \Gamma (\frac{d}{2})}{ 128 (d^2 +d-2)(u+1)^3 \Gamma (\frac{d+3}{2})}.
\end{eqnarray}
The remaining functions $g_j^{(i)}$ with $i \in 2, 3, 4$ and $j \in 1, \ldots 11$ are defined for $i=2$ as
\begin{eqnarray}
g^{(2)}_{1} \hspace{-1mm} &=& \hspace{-1mm}  \frac{ \pi ^2 \, \, 2^{-d-7} \, P_{2,1} \, \Gamma (d-1)}{(u+1)^3 \Gamma \left(\frac{d}{2}+2\right) \Gamma \left(\frac{d}{2}\right)}, 
\\
g^{(2)}_{2} \hspace{-1mm} &=& \hspace{-1mm}  -\frac{ \pi ^2 \, \, 2^{-d-9} P_{2,2} \,  \Gamma (d+3) }{ (d-1) (d+1)^2 u (u+1)^4 (2 u+1) \Gamma \left(\frac{d}{2}+2\right)^2 }, \nonumber \\
\\
g^{(2)}_{3} \hspace{-1mm} &=& \hspace{-1mm}  - \frac{\pi ^2 \, \, 2^{-d-10} \, P_{2,3} \, \Gamma (d+3) }{(d-1) (d+1)^2 u (u+1)^2 (2 u+1)^2 \Gamma \left(\frac{d}{2}+2\right)^2}, \nonumber \\
\\
g^{(2)}_{4} \hspace{-1mm} &=& \hspace{-1mm}  \frac{\pi ^2 \, \, 2^{-d-8} \, P_{2,4} \, \Gamma (d-1) }{u (u+1)^2 (2 u+1)^2 \Gamma \left(\frac{d}{2}+2\right) \Gamma \left(\frac{d}{2}\right)}, 
\\
g^{(2)}_{5} \hspace{-1mm} &=& \hspace{-1mm}  \frac{\pi ^2 \, \, 2^{-d-7} \, P_{2,5} \, \Gamma (d-1)}{(u+1)^2 (2 u+1) \Gamma \left(\frac{d}{2}+1\right) \Gamma \left(\frac{d}{2}+2\right)}, 
\\
g^{(2)}_{6} \hspace{-1mm} &=& \hspace{-1mm}  \frac{3 \pi ^2 \, \, 2^{-d-7} \, P_{2,6} \, \Gamma (d-1)}{(u+1)^2 \Gamma \left(\frac{d}{2}+2\right)^2}, \\
g^{(2)}_{7} \hspace{-1mm} &=& \hspace{-1mm}  - \frac{ \pi ^2 \, \, 2^{-d-7} \, P_{2,7} \, \Gamma (d-1) }{(u+1)^3 (2 u+1) \Gamma \left(\frac{d}{2}+1\right) \Gamma \left(\frac{d}{2}+2\right)}, 
\\
g^{(2)}_{8} \hspace{-1mm} &=& \hspace{-1mm}  -\frac{\pi ^{3/2} \, P_{2,8} \, \Gamma \left(\frac{d-1}{2}\right)}{1024 d (u+1)^5 \Gamma \left(\frac{d}{2}+2\right)}
\\
g^{(2)}_{9} \hspace{-1mm} &=& \hspace{-1mm}  \frac{9 \pi ^{3/2} \, P_{2,9} \, \Gamma \left(\frac{d-1}{2}\right)}{64 d^2 (d+2)^2 (u+1)^4 \Gamma \left(\frac{d}{2}\right)}, 
\\
g^{(2)}_{10} \hspace{-1mm} &=& \hspace{-1mm}  \frac{\pi ^2 \, \, 2^{-d-7} \, P_{2,10} \, \Gamma (d+1)}{(d-1) (u+1)^3 \Gamma \left(\frac{d}{2}+2\right)}, 
\\
g^{(2)}_{11} \hspace{-1mm} &=& \hspace{-1mm}  \frac{3 \pi ^{3/2} \, P_{2,11} \, \Gamma \left(\frac{d-1}{2}\right)}{64 d (d+2) (u+1)},
\end{eqnarray}
where polynomials $P_{2,j}$ with $j \in 1, \ldots, 11$ over $u$ and $d$ have been singled out and are given in the text below. Analogously to the $i=2$ case we get for $i=3$ the following expressions:
\begin{eqnarray}
g^{(3)}_1 \hspace{-1mm} &=& \hspace{-1mm}\frac{\pi ^{3/2} \, P_{3,1} \, \Gamma \left(\frac{d-1}{2}\right)}{32 (d+2) (u+1)},
\\
g^{(3)}_2 \hspace{-1mm} &=& \hspace{-1mm}\frac{\pi ^2 2^{-d-7} \, P_{3,2} \,  \Gamma (d+1)}{(d-1) (u+1)^3 \Gamma \left(\frac{d}{2}+2\right)},
\\
g^{(3)}_3 \hspace{-1mm} &=& \hspace{-1mm}-\frac{3 \pi ^{3/2} u^2 \, P_{3,3} \,  \Gamma \left(\frac{d-1}{2}\right)}{32 d^2 (d+2)^2 (u+1)^4 \Gamma \left(\frac{d}{2}\right)},
\\
g^{(3)}_4 \hspace{-1mm} &=& \hspace{-1mm}-\frac{\pi ^{3/2} \, P_{3,4} \,  \Gamma \left(\frac{d-1}{2}\right)}{512 d (u+1)^5 \Gamma \left(\frac{d}{2}+2\right)},
\\
g^{(3)}_5 \hspace{-1mm} &=& \hspace{-1mm}-\frac{\pi ^2 2^{-d-8} \, P_{3,5} \,  \Gamma (d-1)}{(u+1)^3 (2 u+1) \Gamma \left(\frac{d}{2}+2\right)^2},
\\
g^{(3)}_6 \hspace{-1mm} &=& \hspace{-1mm}\frac{\pi ^2 2^{-d-7} \, P_{3,6} \,  \Gamma (d-1)}{(u+1)^2 \Gamma \left(\frac{d}{2}+2\right)^2},
\\
g^{(3)}_7 \hspace{-1mm} &=& \hspace{-1mm}\frac{\pi ^2 2^{-d-8} \, P_{3,7} \,  \Gamma (d-1)}{(u+1)^2 (2 u+1) \Gamma \left(\frac{d}{2}+2\right)^2},
\\
g^{(3)}_8 \hspace{-1mm} &=& \hspace{-1mm} \frac{\pi ^2 2^{-d-7} \, P_{3,8} \,  \Gamma (d-1)}{u (u+1)^2 (2 u+1)^2 \Gamma \left(\frac{d}{2}+2\right) \Gamma \left(\frac{d}{2}\right)},
\\
g^{(3)}_9 \hspace{-1mm} &=& \hspace{-1mm}-\frac{\pi ^2 2^{-d-9} \, P_{3,9} \,  \Gamma (d+3)}{(d-1) (d+1)^2 u (u+1)^2 (2 u+1)^2 \Gamma \left(\frac{d}{2}+2\right)^2},
\nonumber \\
\\
g^{(3)}_{10} \hspace{-1mm} &=& \hspace{-1mm}-\frac{\pi ^2 2^{-d-8} \, P_{3,10} \,  \Gamma (d+3)}{(d-1) (d+1)^2 u (u+1)^4 (2 u+1) \Gamma \left(\frac{d}{2}+2\right)^2},
\nonumber \\
\\
g^{(3)}_{11} \hspace{-1mm} &=& \hspace{-1mm}\frac{\pi ^{3/2} \, P_{3,11} \,  \Gamma \left(\frac{d-1}{2}\right)} {512 (u+1)^3 \Gamma \left(\frac{d}{2}+2\right)},
\end{eqnarray}
where polynomials $P_{3,j}$ with $j \in 1, \ldots, 11$ over $u$ and $d$ have been singled out and are given in the text below. Analogously to the $i=2,3$ cases we get for $i=4$ the following expressions:
\begin{eqnarray}
g^{(4)}_1  \hspace{-1mm} &=&  \hspace{-1mm}\frac{\pi ^2 2^{-d-7} \, P_{4,1} \, \Gamma (d-1)}{(u+1)^3 \Gamma \left(\frac{d}{2}+2\right) \Gamma \left(\frac{d}{2}\right)}
\\
g^{(4)}_2  \hspace{-1mm} &=&  \hspace{-1mm}-\frac{\pi ^2 2^{-d-5} \, P_{4,2} \, \Gamma (d-1)}{d (d+1) (d+2) u (u+1)^4 (2 u+1) \Gamma \left(\frac{d}{2}\right)^2}
\nonumber \\
\\
g^{(4)}_3  \hspace{-1mm} &=&  \hspace{-1mm}-\frac{\pi ^2 2^{-d-6} \, P_{4,3} \, \Gamma (d-1)}{d (d+1) (d+2) u (u+1)^2 (2 u+1)^2 \Gamma \left(\frac{d}{2}\right)^2}
\nonumber \\
\\
g^{(4)}_4  \hspace{-1mm} &=&  \hspace{-1mm} \frac{\pi ^{3/2} \, P_{4,4} \, \Gamma \left(\frac{d-1}{2}\right)}{1024 u (u+1)^2 (2 u+1)^2 \Gamma \left(\frac{d}{2}+2\right)}
\\
g^{(4)}_5  \hspace{-1mm} &=&  \hspace{-1mm}\frac{\pi ^{3/2} \, P_{4,5} \,  \Gamma \left(\frac{d-1}{2}\right)}{128 d (u+1)^2 (2 u+1) \Gamma \left(\frac{d}{2}+2\right)}
\\
g^{(4)}_6  \hspace{-1mm} &=&  \hspace{-1mm}-\frac{\pi ^2 2^{-d-7} \, P_{4,6} \, \Gamma (d-1)}{(u+1)^2 \Gamma \left(\frac{d}{2}+2\right)^2}
\\
g^{(4)}_7  \hspace{-1mm} &=&  \hspace{-1mm}-\frac{\pi ^{3/2} \, P_{4,7} \, \Gamma \left(\frac{d-1}{2}\right)}{128 d (u+1)^3 (2 u+1) \Gamma \left(\frac{d}{2}+2\right)}
\\
g^{(4)}_8  \hspace{-1mm} &=&  \hspace{-1mm}\frac{\pi ^{3/2} u \, P_{4,8} \, \Gamma \left(\frac{d-1}{2}\right)}{1024 d (u+1)^5 \Gamma \left(\frac{d}{2}+2\right)}
\\
g^{(4)}_9  \hspace{-1mm} &=&  \hspace{-1mm}-\frac{3 \pi ^{3/2} u^2 \, P_{4,9} \, \Gamma \left(\frac{d-1}{2}\right)}{32 d^2 (d+2)^2 (u+1)^4 \Gamma \left(\frac{d}{2}-1\right)}
\\
g^{(4)}_{10}  \hspace{-1mm} &=&  \hspace{-1mm}0
\\
g^{(4)}_{11}  \hspace{-1mm} &=&  \hspace{-1mm}\frac{\pi ^{3/2} \, P_{4,11} \, \Gamma \left(\frac{d-1}{2}\right)}{64 (d+2) (u+1)}
\end{eqnarray}
where polynomials $P_{4,j}$ with $j \in 1, \ldots, 11$ over $u$ and $d$ have been singled out and are given together with $P_{2,j}$ and $P_{3,j}$ for $j \in 1, \ldots, 11$ now in the text below. Let us start with polynomials for $i=2$. We obtain the following expressions: 
\begin{eqnarray}
P_{2,1} \hspace{-1mm} &=& \hspace{-1mm}  d^2 (7 u^2+4u)-3 d (6 u^2+5u) +8 u^2-22 u -30
\nonumber \\ 
\\
P_{2,2} \hspace{-1mm} &=& \hspace{-1mm} 2 d^4 (7 u+6) (u+1)^5-d^3  (u+1)^2 ( 119 u^4 +347u^3
\nonumber \\
&+& \hspace{-1mm} 384u^2 +191u + 33 ) + d^2 (117u^6 + 751u^5
\nonumber \\
&+& \hspace{-1mm} 1733u^4 + 1694u^3+720u^2+74u -19)
\nonumber \\
&+& \hspace{-1mm} d ( 190 u^6  + 82u^5 + 210u^4 +1004u^3 + 1389u^2
\nonumber \\
&+& \hspace{-1mm} 765u+146 ) -8 (2 u+1)(u+1)^2 (5u^3 -19u^2 
\nonumber \\
&+& \hspace{-1mm} 48u +18)
\\
P_{2,3} \hspace{-1mm} &=& \hspace{-1mm}  2 d^5 (6 u+5) (u+1)^5+d^4 (u+1)^2 (19 u^4 -140u^3 
\nonumber \\
&-& \hspace{-1mm} 408 u^2-318 u-73) - d^3 (41 u^6 +16u^5 -251u^4
\nonumber \\
&-& \hspace{-1mm} 750u^3 - 545u^2 - 54u +31) -d^2 (76 u^6+ 3291u^4
\nonumber \\
&+& \hspace{-1mm} 2656 u^3 -370 u^2 -1006u -253) + d ( -28 u^6
\nonumber \\ 
&+& \hspace{-1mm} 1954 u^5 +3757u^4 +2548u^3+76u^2-410u-81)
\nonumber \\
&+& \hspace{-1mm} 2 (2 u+1) (45 u^4 + 636 u^3+ 1236u^2  + 656 u +123)
\nonumber \\
\\
P_{2,4} \hspace{-1mm} &=& \hspace{-1mm} 2 d^4 (6 u+5) (u+1)^5 - d^3 (1 + u)^2 ( 5 u^4+152u^3
\nonumber \\ 
&+& \hspace{-1mm} 352u^2+260u+59) - d^2 ( 31 u^6+258u^5 +606 u^4
\nonumber \\
&+& \hspace{-1mm} 358 u^3+101 u^2 +96 u+38 ) - d (14u^6 -714 u^5
\nonumber \\
&-& \hspace{-1mm} 1419u^4 -1934 u^3 -2166u^2 -1240 u -253) 
\nonumber \\
&-& \hspace{-1mm} 2 (2 u+1) ( 7 u^4-150u^3-252u^2-10u+21) 
\nonumber \\
\\
P_{2,5} \hspace{-1mm} &=& \hspace{-1mm}  d^2(d-2) ( 8 u^3+ 19 u^2+ 14 u +3 ) 
\nonumber \\
&-& \hspace{-1mm} d(d-2)( 122 u^3 + 209 u^2 +102 u +15) 
\nonumber \\
&+& \hspace{-1mm} (d-2)(30 u^3-196 u^2 -188 u -42) 
\\
P_{2,6} \hspace{-1mm} &=& \hspace{-1mm} d^2 (d+2)(1 + 3 u + 2 u^2)
\nonumber \\
&-& \hspace{-1mm} d (d+2)(3 + 4 u + 5 u^2) - 8(d+2)u
\\
P_{2,7} \hspace{-1mm} &=& \hspace{-1mm} d^2 (d-2)(8 u+3) (u+1)^3 
\nonumber \\
&-& \hspace{-1mm} 2 d (d-2)(u+1) (37 u^3 + 56 u^2 + 24 u + 3 ) 
\nonumber \\ 
&+& \hspace{-1mm} 8(d-2) (2 u+1) (2 u^3 - 6 u^2 - 10 u - 3 )
\\
P_{2,8} \hspace{-1mm} &=& \hspace{-1mm} 2 d^2 (u+1)^2 (3 u^2+5u+6) + d (43 u^4 + 155 u^3
\nonumber \\ 
&+& \hspace{-1mm}  155 u^2 + 49u) - 14 u^4 - 6 u^3 + 110 u^2 + 150 u + 48
\nonumber \\
\\
P_{2,9} \hspace{-1mm} &=& \hspace{-1mm} u^2 \left[ d(5 u+1)-2 u+6 \right]
\\
P_{2,10} \hspace{-1mm} &=& \hspace{-1mm} d (10 u+7)
\\
P_{2,11} \hspace{-1mm} &=& \hspace{-1mm} (3 d+2 u)
\end{eqnarray}

\begin{eqnarray}
P_{3,1} \hspace{-1mm} &=& \hspace{-1mm} 5 u+1,
\\
P_{3,2} \hspace{-1mm} &=& \hspace{-1mm} d (2 u+1),
\\
P_{3,3} \hspace{-1mm} &=& \hspace{-1mm} d (5 u+1)-14 u-6,
\\
P_{3,4} \hspace{-1mm} &=& \hspace{-1mm} d^2 (u+2) (u+1)^3-d u (13 u^3+51 u^2+53 u +17)
\nonumber \\
&+& \hspace{-1mm} 2(u+1)(19 u^3+54 u^2+31 u+4),
\\
P_{3,5} \hspace{-1mm} &=& \hspace{-1mm} d^2(d-2)(d+2)\left[ \, (4u+1)(u+1)^3 \right. 
\nonumber \\
&-& \hspace{-1mm} 2 d (u+1)(4 u^3+11 u^2+6 u+1)
\nonumber \\
&-& \hspace{-1mm} \left. 8 (2 u+1)(10 u^3+15 u^2+8 u+1) \, \right],
\\
P_{3,6} \hspace{-1mm} &=& \hspace{-1mm} d^2(d+2)(u+1)(2u+1)- 2d(d+2)(5u+1)
\nonumber \\
&-&  \hspace{-1mm} 4(d+2)u(7 u+3),
\\
P_{3,7} \hspace{-1mm} &=& \hspace{-1mm} ( d^2-4 ) \left[ d (u+1) -15 u -7 \right] \,  \left[ 4 (d+4) u^2 \right.
\nonumber \\
&+& \left. (5 d+14) u+d+2\right],
\\
P_{3,8} \hspace{-1mm} &=& \hspace{-1mm} d^4 (4u+3)(u+1)^5 + d^3 (u+1)^2 (11 u^4 - 13 u^3
\nonumber \\
&-& \hspace{-1mm} 80 u^2-73 u-19) - d^2 (15 u^6+304 u^5 + 905 u^4
\nonumber \\
&+& \hspace{-1mm} 904 u^3 + 379 u^2+60 u+1) - d (22 u^6 - 106 u^5
\nonumber \\
&-& \hspace{-1mm} 771 u^4-1396 u^3-1104 u^2 - 410u - 59)
\nonumber \\
&+& \hspace{-1mm} 2(2 u+1)(u+1)^2(67 u^2+22u -5)
\\
P_{3,9} \hspace{-1mm} &=& \hspace{-1mm} d^5 (4 u+3) (u+1)^5 + d^4 (u+1)^2 (19 u^4 +u^3
\nonumber \\
&-& \hspace{-1mm} 76 u^2 -76 u-20) + d^3 (7 u^6-268 u^5 - 938 u^4
\nonumber \\
&-& \hspace{-1mm} 1002 u^3-449 u^2-82 u-4) - d^2 (52 u^6+170 u^5 
\nonumber \\
&+& \hspace{-1mm} 139 u^4 - 453 u^3-613 u^2 - 253 u-34)- d (44 u^6
\nonumber \\
&+& \hspace{-1mm} 210 u^5+129 u^4-134 u^3-176 u^2-76 u-13)
\nonumber \\
&+& \hspace{-1mm} 2 (2 u+1) (175 u^4+572 u^3 + 610 u^2+276 u+47),
\nonumber \\
& &
\\
P_{3,10} \hspace{-1mm} &=& \hspace{-1mm} 2 d^4 (u+1)^6 - d^3 (u+1)^2 (6 u^4+22 u^3+35 u^2
\nonumber \\
&+& \hspace{-1mm} 28 u+8) - d^2 (2 u+1) (54 u^5+129 u^4+78 u^3
\nonumber \\
&+& \hspace{-1mm} 4 u^2-27 u-12) +d (u+1)(108 u^5+466 u^4 
\nonumber \\
&+& \hspace{-1mm} 700 u^3 + 474 u^2+149 u+18) +4 (2 u+1) \times
\nonumber \\
& & (u+1)^2 (26 u^3 - 24u^2 - 41 u-13),
\\
P_{3,11} \hspace{-1mm} &=& \hspace{-1mm} u \left[ (d-2) (d+16) u+4 (3 d-8) \right] ,
\end{eqnarray}

\begin{figure*}
\begin{center}\includegraphics[width=17cm]{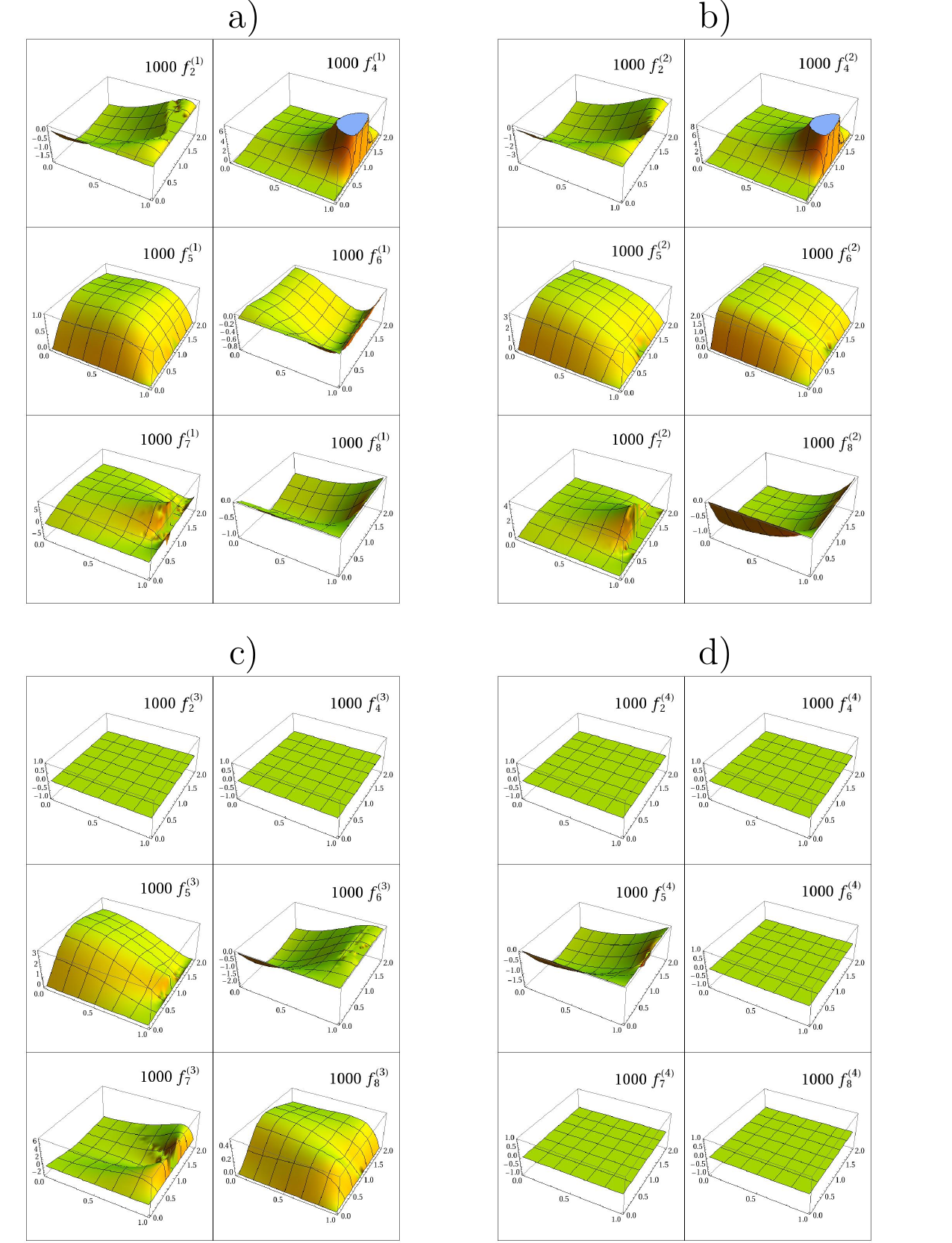}\end{center}
\caption{ (Color online) Functions $f^{(i)}_l$ for $l \in 2,4,5,6,7,8$ and $i\in 1,2,3,4$ shown for $u$ corresponding to a value typical of the $A=-1$ model. Values of $f^{(0)}$ have already been determined within the scope of the $A=0$ model in Ref.\,\cite{Jurcisin2014}. Subfigure a) depicts coefficients $f^{(1)}_l$, subfigure b) shows $f^{(2)}_l$, subfigure c) shows $f^{(i)}_l$ and subfigure d) shows $f^{(4)}_l$. Note that only the fifth graph actually contributes terms of order $A^4$ to the final expression for $z_{21}^{(2)}$ since all the other graphs contribute zero coefficients $f^{(4)}_l$ as shown by flat planes in the corresponding figures. All dependences are plotted in region of $x \in \langle 0, 1\rangle$ and $z \in \langle 0,2 \rangle.$ \label{fig8} }
\end{figure*}

\begin{eqnarray}
P_{4,1}  \hspace{-1mm} &=&  \hspace{-1mm} (d-2)(4 u^2+5u)-2,
\\
P_{4,2}  \hspace{-1mm} &=&  \hspace{-1mm} d^3 (u+1)^4 (13 u^2+15u+1) - d^2 (u+1)(109 u^5 
\nonumber \\
&+&  \hspace{-1mm} 366 u^4 + 505 u^3 + 335 u^2 + 91 u + 5) +d (54 u^6
\nonumber \\ 
&+&  \hspace{-1mm} 970 u^5+2150 u^4+2110 u^3+1007 u^2+205 u +10)
\nonumber \\ 
&+&  \hspace{-1mm} 8 (2 u+1) (u+1)^2 (11 u^3-39 u^2-19 u-1),
\\
P_{4,3}  \hspace{-1mm} &=&  \hspace{-1mm} d^4 (u+1)^4 (19 u^2+16u+1) + d^3 (35 u^4-254 u^3
\nonumber \\
&-&  \hspace{-1mm} 422 u^2-160 u-9) (u+1)^2 -d^2 (28 u^6+454 u^5
\nonumber \\
&-&  \hspace{-1mm} 265 u^4-1534 u^3-1164 u^2-288 u-15) - d (44 u^6
\nonumber \\
&+&  \hspace{-1mm} 1630 u^5+2699 u^4+2272 u^3+1004 u^2 + 190 u+9)
\nonumber \\
&+&  \hspace{-1mm} 2 (2 u+1) (485 u^4+1084 u^3+664 u^2+152 u+7 ),
\nonumber \\
& &
\\
P_{4,4}  \hspace{-1mm} &=&  \hspace{-1mm} d^3 (u (19 u+16)+1) (u+1)^4 - d^2 (u+1)^2 (3 u^4
\nonumber \\
&+&  \hspace{-1mm} 272 u^3+389 u^2+144 u+8) - d (22 u^6+430 u^5
\nonumber \\
&-&  \hspace{-1mm} 175 u^4-1290 u^3-1018 u^2-260 u-13)
\nonumber \\
&+&  \hspace{-1mm} 2 (2 u+1) (113 u^4+226 u^3+32 u^2-18 u-1),
\nonumber \\
\\
P_{4,5}  \hspace{-1mm} &=&  \hspace{-1mm}  (d-2) u (5 u+2)  \left[ \, (d-15) u+d-7 \, \right],
\\
P_{4,6}  \hspace{-1mm} &=&  \hspace{-1mm} (d+2) (5 u+1) (d+2 u),
\\
P_{4,7}  \hspace{-1mm} &=&  \hspace{-1mm} (d-2) u \left[ d (u+1)^2 (5 u+2)-4 (2 u+1) (4 u^2 \right.
\nonumber \\
&+&  \hspace{-1mm} \left. 7u+2) \right],
\\
P_{4,8}  \hspace{-1mm} &=&  \hspace{-1mm} d (17 u^3+49u^2+49u+15)
\nonumber \\
&-&  \hspace{-1mm} 2 (u+1)^2 (17 u+15),
\\
P_{4,9}  \hspace{-1mm} &=&  \hspace{-1mm} 5 u+1,
\\
P_{4,10}  \hspace{-1mm} &=&  \hspace{-1mm} 0,
\\
P_{4,11}  \hspace{-1mm} &=&  \hspace{-1mm} 5 u+1.
\end{eqnarray}
Let us turn our attention to the remaining seven diagrams of Fig.\,\ref{fig3}. Instead of calculating the graphs using the previously discussed technique of Ref.\,\cite{VasilevBook}, we have employed the derivative technique outlined by authors of Ref.\,\cite{Adzhemyan2005} as it allows an easy algorithmic approach.  In analogy to Eq.\.(\ref{StructureGraph1}), for each diagram $l \in 2, \ldots 8$ we may explicitly determine every coefficient of the resulting polynomial over $A$ separately for each graph. The corresponding decomposition reads now:
\begin{equation}
B^{(\rho)}_l = 16 u \int\limits_{0}^{1} \hspace{-1mm} dx (1-x^2)^{(d-1)/2} \hspace{-1mm} \int\limits_{1}^{\infty} \hspace{-1mm} \frac{dz}{z} \sum_{i=0}^{4} A^i \, f^{(i)}_l(x,z) \label{SevenGraphs}
\end{equation}
with $l \in 2, \ldots 8$. Notice that contrary to the expression (\ref{StructureGraph1}), here integrations over variables $x \in \langle -1,1 \rangle$ with $x ={\mathbf k}.{\mathbf q}/|k|/|q|$ and $z \geq 1$ are singled out. Each of the functions $f^{(i)}_l(x,z,u,d)$ is a rational function over $x,z,u$ and $d$. Since for $l \in 2, 4, 5, 6, 7, 8$  the corresponding expressions are lengthy and require huge amount of space we do not show the explicit form of their corresponding $f^{(i)}_l$ functions. Instead, as an example, we give now the corresponding expressions only for the third two-loop diagram $\Gamma^{(2)}_3$ of Fig.\,\ref{fig3} and present the remaining functions $f^{(i)}_l$ for $l \in 2, 4, 5, 6, 7, 8$ graphically in Fig.\,\ref{fig7}. Since $A=0$ case has been completely solved in Ref.\,\cite{Jurcisin2014}, we now present only $f^{(1)}_3(x,z,u,d)$ and $f^{(2)}_3(x,z,u,d)$. As discussed in the main body of the article, for the present diagram $f^{(3)}_3(x,z,u,d)=f^{(4)}_3(x,z,u,d)=0$ because of the structure of $A$ polynomials. Thus, we get:
\begin{widetext}
\begin{eqnarray}
f^{(1)}_3(x,z,u,d) &=& \frac{(-2+d) (d-2 u+d u)  z \left(1+z^2\right) \left(-\left(1+z^2\right)^2+x^2 \left(3-2 z^2+3 z^4\right)\right)}{16 (-1+d) d (2+d) (1+u)^2 \left(-1+x z-z^2\right) \left(-1+2 x z-z^2\right) \left(1+x z+z^2\right) \left(1+2 x z+z^2\right)},
\\
f^{(2)}_3(x,z,u,d) &=& \frac{(-2+d) (1+3 u)  z \left(1+z^2\right) \left(-\left(1+z^2\right)^2+x^2 \left(3-2 z^2+3 z^4\right)\right)}{16 (-1+d) d (2+d) (1+u)^2 \left(-1+x z-z^2\right) \left(-1+2 x z-z^2\right) \left(1+x z+z^2\right) \left(1+2 x z+z^2\right)}.
\end{eqnarray}
\end{widetext}
As already discussed, graph $\Gamma^{(2)}_3$ contains only two $V_{ijl}$ type of vertices and its corresponding functions $f^{(3)}_l(x,z,u,d)$ and $f^{(4)}_l(x,z,u,d)$, which correspond to polynomial coefficients in front of $A^3$ and $A^4$ respectively, are both zero. As already discussed, for the remaining graphs $\Gamma^{(2)}_l$ with $l \in 2, 4, 5, 6, 7, 8$ the $f^{(i)}_l(x,z,u,d)$ functions with $i \in 0. \ldots 4$ are lengthy and require huge amount of space and we shall only present them graphically via Fig.\,\ref{fig7} for $u=1$ which demonstrates their usual shape as used in the actual calculations of turbulent Prandtl number via Eq.\,(\ref{InversePrandtl}).


\begin{references}


\bibitem{Jurcisin2014}
E.~Jur\v{c}i\v{s}inov\'a, M.~ Jur\v{c}i\v{s}in, and P.~ Zalom, Phys. Rev. E  \textbf{89}, 043023 (2014).

\bibitem{Yoshizawa}
A.~Yoshizawa, S.~I.~Itoh, and K.~Itoh, \textit{Plasma and Fluid Turbulence: Theory and Modelling} (IoP, Bristol and Philadelphia, 2003).

\bibitem{Biskamp}
D.~Biskamp, \textit{Magnetohydrodynamic Turbulence} (CUP,Cambridge, 2003).

\bibitem{MoninBook}
A.~S.~Monin and A.~M.~Yaglom, \textit{Statistical Fluid Mechanics} (MIT Press, Cambridge, MA, 1975), Vol. 2.

\bibitem{McComb}
W.~D.~McComb, \textit{The Physics of Fluid Turbulence} (Clarendon, Oxford, 1990).

\bibitem{Shraiman}
B.~I.~Shraiman and E.~D.~Siggia, Review Nature \textbf{405}, 639-646 (2000)

\bibitem{Coulson}
J.~M.~Coulson, J.~F.~Richardson, \textit{Chemical Engineering} (Elsevier, 1999), Vol.1.

\bibitem{Chua}
L.~P.~Chua and R.~A.~Antonia, Int. J.~Heat Mass Transf. \textbf{33}, 331 (1990).

\bibitem{Chang}
L.~P.~Chang and E.~A.~Cowen, J.~ Eng. Mech. \textbf{128}, 1082 (2002).

\bibitem{VasilevBook}
A.~N.~Vasil’ev, \textit{Quantum-Field Renormalization Group in the Theory of Critical Phenomena and Stochastic Dynamics} (Chapman \& Hall/CRC, Boca Raton, 2004).

\bibitem{AdzhemyanBook} 
L.~Ts.~Adzhemyan, N.~V.~Antonov, and A.~N.Vasil’ev, \textit{The Field Theoretic Renormalization Group in Fully Developed Turbulence} (Gordon \& Breach, London, 1999).

\bibitem{Antonov2015}
N.~V.~Antonov and N.~M.~Gulitskiy, Phys. Rev. E  \textbf{91}, 013002 (2015)

\bibitem{Antonov2015a}
N.~V.~Antonov and N.~M.~Gulitskiy, Phys. Rev. E  \textbf{92}, 043018 (2015)

\bibitem{Jurcisin2016}
E.~Jur\v{c}i\v{s}inov\'a, M.~Jur\v{c}i\v{s}in, and R.~Remeck\'y, Phys. Rev. E  \textbf{93}, 033106 (2016).

\bibitem{Adzhemyan2005}
L.~Ts.~Adzhemyan, J.~Honkonen, T.~L.~Kim, and L.~Sladkoff, Phys. Rev. E  \textbf{71}, 056311 (2005).

\bibitem{Arponen2009}
H.~Arponen, Phys. Rev. E \textbf{79}, 056303 (2009).

\bibitem{Antonov2003}
N.~V.~Antonov, Phys. Rev. E \textbf{68}, 046306 (2003).

\bibitem{Zinn}
J.~Zinn-Justin, Quantum Field Theory and Critical Phenomena (Clarendon, Oxford, 1989).

\bibitem{Adzhemyan1983}
L.~Ts.~Adzhemyan, A.~N.~Vasil’ev, and Yu.~M.~ Pis’mak, Theor. Math. Phys. \textbf{57}, 1131 (1983).

\bibitem{Adzhemyan2003}
L.~Ts.~Adzhemyan, N.~V.~Antonov, M.~V.~Kompaniets, and A.~N.~Vasil’ev, Int. J. Mod. Phys. B \textbf{17}, 2137 (2003).

\bibitem{Adzhemyan2003a}
L.~Ts.~Adzhemyan, J.~Honkonen, M.~V.~Kompaniets, and A.~N.~Vasil’ev, Phys. Rev. E  \textbf{68}, 055302(R) (2003).

\bibitem{Adzhemyan1988}
L.~Ts.~Adzhemyan, A.~N.~Vasil’ev and M.~Gnatich, Theor. Math. Phys. \textbf{74}, 180-191 (1988).

\bibitem{Adzhemyan1996}
L.~Ts.~Adzhemyan, N.~V.~Antonov, and A.~N.~Vasil’ev, Usp. Fiz. Nauk \textbf{166}, 1257 (1996) [Phys. Usp. 39, 1193 (1996)].

\bibitem{Adzhemyan2005double}
L.~Ts.~Adzhemyan, J.~Honkonen, M.~V.~Kompaniets, and A.~N.~Vasiliev, Phys. Rev. \textbf{E} 71, 036305 (2005).

\bibitem{Adzhemyan2006double}
L.~Ts.~Adzhemyan, J.~Honkonen, T.~L.~Kim, M.~V.~Kompaniets, L.~Sladkoff, and A.~N.~Vasil'ev, J. Phys. A: Math. Gen. \textbf{39} 7789 (2006).

\bibitem{Adzhemyan1983a}
L.~Ts.~Adzhemyan, A.~N.~Vasil’ev, and M.~Gnatich, Theor. Math. Phys. \textbf{58}, (1983).

\bibitem{Adzhemyan1998}
L.~Ts.~Adzhemyan, N.~V.~Antonov and A.~N.~Vasiliev, Phys. Rev. E  \textbf{58}, 1823 (1998)

\bibitem{Adzhemyan2001}
L.~Ts.~Adzhemyan, N.~V.~Antonov, V.~A.~Barinov, Yu.~S.~Kabrits, and A.~N.~Vasiliev, Phys. Rev. \textbf{E} 63, 025303(R) (2001); L.~Ts.~Adzhemyan, N.~V.~Antonov, V.~A.~Barinov, Yu.~S.~Kabrits, and A.~N.~Vasil’ev Phys. Rev. \textbf{E} 64, 019901 (2001).

\bibitem{Novikov2003}
S.~V.~Novikov, Theor. Math. Phys. 136, 936 (2003)

\bibitem{Pagani2015}
C.~Pagani, Phys. Rev. \textbf{E} 92, 033016 (2015).

\bibitem{Adzhemyan1985}
L.~Ts.~Adzhemyan, A.~N.~Vasil’ev, and M.~Gnatich, Theor. Math. Phys. \textbf{64}, 777 (1985).

\bibitem{Adzhemyan1987}
L.~Ts.~Adzhemyan, A.~N.~Vasiliev, and M.~Hnatich, Theor. Math. Phys. \textbf{72}, 940 (1987).

\bibitem{Adzhemyan2013}
L.~Ts.~Adzhemyan, N.~V.~Antonov, P.~B.~Goldin, and M.~V.~Kompaniets, J. Phys. A: Math. Theor. \textbf{46}, 135002 (2013); 
N.~V.~Antonov and N.~M.~Gulitskiy, Theor. Math. Phys. \textbf{176}, 851 (2013).

\bibitem{Novikov2006}
S.~V.~Novikov, J. Phys. A: Math. Gen. \textbf{39} 8133 (2006).

\bibitem{Arponen2010}
H. Arponen, Phys. Rev. \textbf{E} 81, 036325 (2010).

\bibitem{Martin}
P.~C.~Martin, E.~D.~Siggia, and H.~A.~Rose, Phys. Rev. A \textbf{8}, 423 (1973); C.~De~Dominicis, J. Phys. (Paris), Colloq. 37, C1-247 (1976); H.~K.~Janssen, Z. Phys. B \textbf{23}, 377 (1976); R.~Bausch, H.~K.~Janssen, and H.~Wagner, ibid. \textbf{24}, 113 (1976).

\bibitem{Collins} 
J.~C.~Collins, \textit{Renormalization: An Introduction to Renormalization, the Renormalization Group and the Operator-Product Expansion}

\end{references}
\end{document}